\UseRawInputEncoding
\documentclass[%
reprint,
 amsmath,amssymb,
 aps,
floatfix,
]{revtex4-2}

\usepackage{graphicx}
\usepackage{dcolumn}
\usepackage{bm}
\usepackage{algorithmic}
\usepackage{graphicx}
\usepackage{textcomp}
\usepackage{placeins}
\usepackage[dvipsnames]{xcolor}
\usepackage{float}
\usepackage{esvect}

\usepackage{subcaption}
\begin{document}

\preprint{APS/123-QED}

\newcommand{\JC}[1]{{\textcolor{black}{#1}}}
\newcommand{\LD}[1]{{\textcolor{black}{#1}}}
\newcommand{\GR}[1]{{\textcolor{ForestGreen}{#1}}}

\title{Electron-Electron Interactions in Device Simulation via Non-equilibrium Green's Functions and the GW Approximation}

\author{Leonard Deuschle}
 \email{dleonard@iis.ee.ethz.ch}
\author{Jiang Cao}%
\author{Alexandros Nikolaos Ziogas}%
\author{Anders Winka}%
\author{Alexander Maeder}%
\author{Nicolas Vetsch}
\author{Mathieu Luisier}

\affiliation{%
 Integrated Systems Laboratory,
ETH Zurich,
Switzerland
}%





\begin{abstract}
The continuous scaling of metal-oxide-semiconductor field-effect transistors (MOSFETs) has led to device geometries where charged carriers are increasingly confined to ever smaller channel cross sections. This development is associated with reduced screening of long-range Coulomb interactions. To accurately predict the behavior of such ultra-scaled devices, electron-electron (e-e) interactions must be explicitly incorporated in their quantum transport simulation. In this paper, we present an \textit{ab initio} atomistic simulation framework based on density functional theory, the non-equilibrium Green's function formalism, and  the self-consistent GW approximation to perform this task. The implemented method is first validated with a carbon nanotube test structure before being applied to calculate the transfer characteristics of a silicon nanowire MOSFET in a gate-all-around configuration. As a consequence of e-e scattering, the energy and spatial distribution of the carrier and current densities both significantly change, while the on-current of the transistor decreases owing to the Coulomb repulsion between the electrons. Furthermore, we demonstrate how the resulting bandgap modulation of the nanowire channel as a function of the gate-to-source voltage could potentially improve the device performance. To the best of our knowledge, this study is the first one reporting large-scale atomistic quantum transport simulations of nano-devices under non-equilibrium conditions and in the presence of e-e interactions within the GW approximation.
\end{abstract}

\maketitle


\section{\label{sec:intro}Introduction}

In recent years, leading semiconductor chip manufacturing companies have started the transition from triple-gate Fin field-effect transistors (FinFET) to gate-all-around (GAA) architectures. Stacked GAA nano-X FETs (NXFETs), where X stands for sheet~\cite{7998183, Mukesh2022}, fork~\cite{Weckx2019}, ribbon~\cite{Huang2020}, or wire~\cite{Mertens2017} have emerged as viable transistor technologies in future nodes due to their excellent electrostatic properties and potentially high on-state current densities. Research has also been conducted on materials replacing silicon such as carbon nanotubes (CNT), which are projected to improve energy efficiency by one order of magnitude compared to today's silicon-based logic switches~\cite{Hills2018, doi:10.1021/nn406301r}. In fact, a modern microprocessor implementation that is based entirely on CNTFETs has been demonstrated in \cite{Hills2019}. 

Due to the small cross section of the NXFETs and CNTFETs , the electron population is heavily confined in ultra-small volumes. Therefore, increased electron-electron (e-e) interactions are expected to occur simultaneously~\cite{Fei2013}. This many-body effect can potentially have a strong impact on device operation. It has been shown to induce Auger recombination processes~\cite{auger, Xia2024}, avalanche breakdowns~\cite{avalanche,iedm23}, and re-distributions of the electron population and electrical current with respect to their energy~\cite{renorm}. To accurately predict the performance of the aforementioned ultra-scaled devices, a modeling framework that offers a quantum mechanical description of their behavior, captures atomistic effects, operates at the \textit{ab initio} level, and includes e-e scattering is imperatively needed. Otherwise, tunneling currents would not be accounted for, the bandstructure of the underlying materials would not be correctly reproduced, the influence of isolated defects or impurities would be missed, and the carrier population would not be able to thermalize in regions with high electric fields. 

The non-equilibrium Green's function (NEGF) method~\cite{Datta2000} is a powerful tool to fulfill these requirements. It allows for the simulation of quantum transport in atomically-resolved devices driven out of equilibrium by external stimuli (voltage, temperature, light)~\cite{Lee2023}. One of the key ingredients of the NEGF formalism, its Hamiltonian matrix $\mathbf{H}$, can be expressed within an \textit{ab initio} basis, thus capturing the ground-state electronic properties of nanowires and nanotubes very accurately. Density functional theory (DFT) is the method of choice for that purpose. It very often relies on a plane-wave (PW) expansion of the system's wavefunctions. This PW basis can be directly combined with NEGF~\cite{PhysRevB.102.045410}. Alternatively, it can be transformed into a set of maximally localized Wannier functions (MLWFs)~\cite{wannier90}, which give rise to a tight-binding-like Hamiltonian matrix, i.e., few basis elements per ``atom'' are retained and the atom-to-atom interaction distances are very short. 

Moreover, NEGF lends itself naturally to the inclusion of many-body interactions such as electron-electron scattering through dedicated self-energies. Here, we implement the GW approximation to e-e scattering~\cite{Reining2017}. GW has become a popular approach to predict electronic excitations in chemical compounds and materials. In particular, it overcomes some of the most notorious deficiencies of DFT, namely self-interaction errors and Kohn-Sham bandgap underestimations~\cite{Golze2019, PhysRevB.36.6497}. Because it also encompasses the desired carrier redistribution effects or avalanche processes, the GW approximation is ideally suited to investigate the targeted NXFETs and to compute their ``current vs. voltage'' characteristics. 

Due to the interdependence of the Green's function $G$ and the screened interaction $W$ in the GW method, their governing equations must be solved iteratively until self-consistency is achieved. However, due to the high computational cost of this process, the calculations are commonly stopped after the first iteration. This approximation is known as $\mathrm{G_0W_0}$ and is considered one of the most accurate methods to determine the spectral properties of solids~\cite{Aulbur2000}. On the other hand, $\mathrm{G_0W_0}$ violates physical conservation laws \cite{Stan2009}. Especially, in non-equilibrium quantum transport simulations, current conservation is not ensured by the $\mathrm{G_0W_0}$ scheme, hence preventing the study of nano-devices. A full self-consistent treatment between $G$ and $W$ is required in this case. Although the Coulomb interaction $V$ can be reduced to a two-index function to save computational and memory resources, DFT + NEGF calculations with a fully self-consistent treatment of the GW terms have been restricted to small molecules made of tens of atoms~\cite{PhysRevB.77.115333}. 

Here, we address this issue and demonstrate an advanced quantum transport simulation environment based on DFT + NEGF + self-consistent GW (scGW) that is capable of treating thousands of atoms and ultra-scaled device geometries. The structure of the paper is as follows: In Section~\ref{sec:Method}, we describe the mathematical foundation of our DFT + NEGF + scGW model with an emphasis on the boundary conditions for $G$ and $W$, the introduction of the latter representing a major innovation of this work. In Section~\ref{sec:Results}, we validate our implementation of the method by using single-walled carbon nanotubes as examples and by testing the influence of the Coulomb interaction range. Next, we simulate the transport properties of a silicon GAA nanowire FET (NWFET). The impact of electron-electron scattering is revealed by comparing the transfer characteristics of this transistor in the presence of e-e scattering and in the commonly adopted quasi-ballistic limit of transport. Importantly, the electronic and energy currents are satisfied throughout the NWFET, as expected. Finally, Section~\ref{sec:Conclusions} summarizes the main contributions of this work.

\section{Method}\label{sec:Method}


\begin{figure*}[ht]
    \centering
        \includegraphics[width=0.8\textwidth]{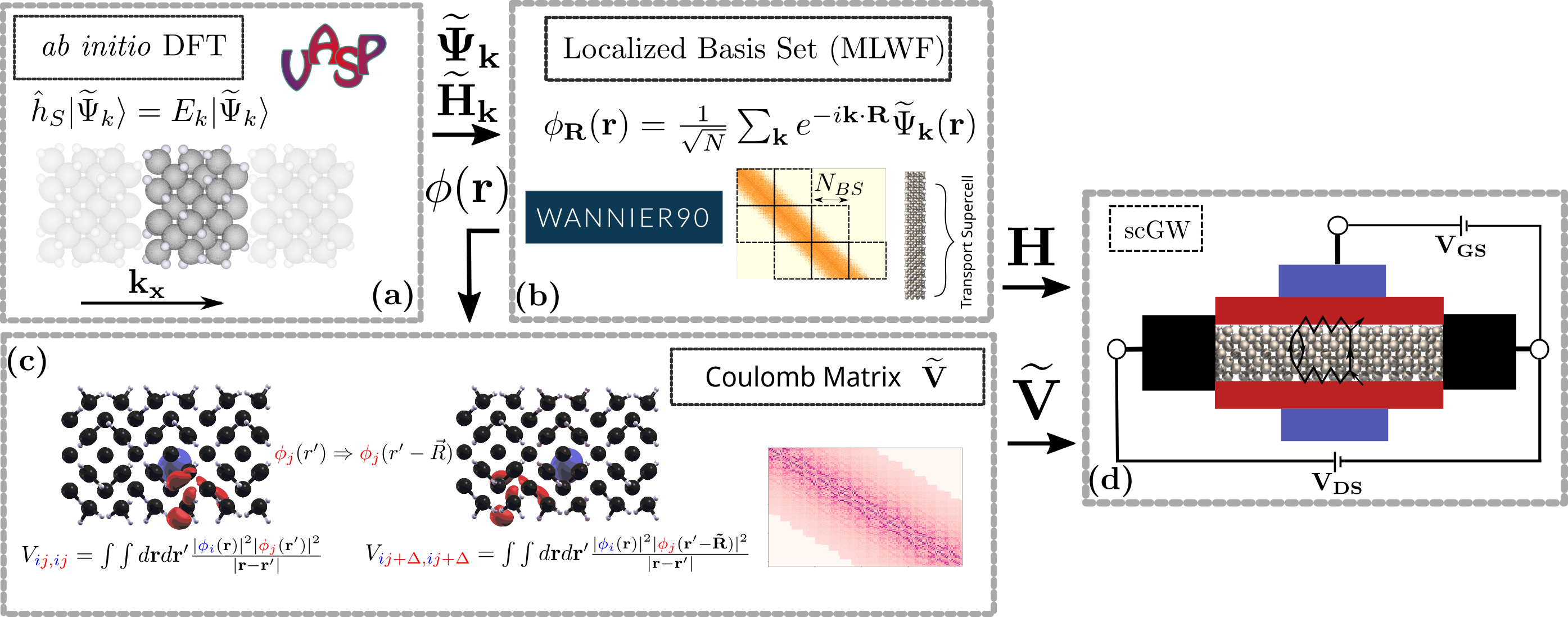}
              \caption{Illustration of the DFT+NEGF+scGW simulation workflow with a silicon nanowire structure as an example. \textbf{(a)} Electronic structure calculation with DFT using the VASP \cite{vasp} code. The primitive nanowire unit cell of width  $\Delta = 0.547$ nm and its periodic images along the transport direction ($x$) are displayed. By diagonalizing the corresponding Hamiltonian $\hat{h}_k$, we obtain the energies $E_k$ and the wave functions $\widetilde{\Psi}_k$, where $k$ is the wave vector representing the periodicity along $x$ \textbf{(b)} Transformation of the $\widetilde{\Psi}_k$ into a set of maximally localized Wannier functions $\phi_R(\mathbf{r})$ with the wannier90 tool \cite{wannier90}. The unit cell Hamiltonian $\widetilde{H}_k$ is re-expressed in terms of a real-space basis and upscaled to the device dimensions, which produces $\mathbf{H}$. \textbf{(c)} Construction of the bare Coulomb matrix $\widetilde{\mathbf{V}}$ with the $\phi(\mathbf{r})$ as inputs. Its diagonal and off-diagonal blocks are calculated by shifting the wavefunctions along the nanowire axis by $\Delta$, the unit cell width. The required operations takes advantage of the GPAW package \cite{Mortensen2024}. \textbf{(d)} DFT+NEGF+scGW simulation of the considered silicon GAA NWFET with the device Hamiltonian $\mathbf{H}$ and $\widetilde{\mathbf{V}}$ as inputs. The drain-to-source $V_{DS}$ and gate-to-source $V_{GS}$ voltages can be varied to obtain the ``current vs. voltage'' characteristics of this device.}
            \label{fig:flowchart}
\end{figure*}

This section presents the developed simulation approach whose general organization is provided in Fig.~\ref{fig:flowchart}. The process starts by identifying a representative unit cell for the device of interest and calculating its electronic properties with the help of plane-wave \textit{ab initio} DFT (Fig.~\ref{fig:flowchart}(a)). The results are then transformed into a localized basis set (Fig.~\ref{fig:flowchart}(b)), which is leveraged to create the Hamiltonian ($\mathrm{\mathbf{H}}$) and the bare Coulomb ($\mathrm{\mathbf{\widetilde{V}}}$) matrices of the system (Fig.~\ref{fig:flowchart}(c)). Both quantities serve as ingredients to the equations at the core of the DFT+NEGF+scGW approach. A particular attention is put on the treatment of the open boundary conditions inherent to quantum transport problems and on the required conditions to ensure electronic and energy current conservation. All these steps are discussed in great detail in the following subsections. Note that in this work, we discuss the treatment of 3-D geometries without transverse electron momentum. All our examples are configured such that electrons are injected from the source contact on the left and flow along a single transport direction labeled $x$ in Fig.~\ref{fig:flowchart}(d). Subsequently, these electrons are extracted at the drain contact on the right side.

\subsection{Electronic Structure and Coulomb Matrix Calculation \label{subsec:DFT}}

\subsubsection{Device Hamiltonian Matrix $\mathbf{H}$}

The \textit{ab initio} plane-wave code VASP~\cite{vasp} is used to calculate the plane-wave Hamiltonian $\hat{h}_k$ and the associated Kohn-Sham energies $E_k$ corresponding to a representative device unit cell, assuming that the transport direction ($x$) is periodic. As exchange-correlation functional, the general gradient approximation (GGA) of Perdew, Burke, and Ernzerhof (PBE) is employed \cite{Perdew1996}. The periodic Bloch orbitals are then transformed into MLWFs with the wannier90 tool \cite{wannier90}. This operation produces blocks connecting Wannier centers within the device unit cell and between adjacent cells. The obtained blocks can be rearranged and scaled up to match the device dimensions and form its Hamiltonian matrix $\mathbf{H}$. This matrix typically has a block-tridiagonal (BT) sparsity pattern and is of size $N_{AO}\times N_{AO}$, $N_{AO}$ being the total number of atomic orbitals (Wannier functions) in the structure. The diagonal blocks of $\mathbf{H}$ contain the on-site elements describing the interactions between all orbitals belonging to the same unit cell. The upper- and lower off-diagonal blocks couple unit cell $i$ to its next ($i+1$) and previous ($i-1$) neighbors, respectively. Due to the homogeneous nature of the device geometries discussed in this work, the size of the diagonal and off-diagonal blocks does not vary along the $x$ axis and is equal to $d\times d$.

\subsubsection{Bare Coulomb Matrix $\mathrm{\mathbf{\widetilde{V}}}$}\label{subsec:CM}
The real-space representation of the MLWFs can be utilized to obtain the entries of the bare Coulomb matrix $\mathrm{\mathbf{V}}$. In general, two-body unscreened Coulomb interactions have the following four-index form~\cite{PhysRevB.77.115333}:
\begin{equation}\label{eq:4index_coul}
V_{ij,kl} = \frac{q^2}{4\pi\epsilon_0}\int \int d\mathbf{r}d\mathbf{r'} \frac{\phi_i(\mathbf{r})^{*}\phi_j(\mathbf{r'})^{*}\phi_k(\mathbf{r})\phi_l(\mathbf{r'})}{|\mathbf{r} - \mathbf{r'}|}.
\end{equation}
In Eq.~(\ref{eq:Coulomb_matrix}), $q$ is the elementary charge, $\epsilon_0$ the vacuum permittivity, and $\phi_i(r)$ denotes a real-space MLWF centered at position $\mathbf{R}_i$. Since calculating all $V_{ij,kl}$ elements becomes computationally very intensive as the number of atoms increases, we adopt the method of Ref.~\cite{PhysRevB.77.115333} and truncate Eq.~(\ref{eq:4index_coul}) to two indices, which leaves us with
\begin{align}\label{eq:Coulomb_matrix}
    \widetilde{V}_{ij} &= \frac{q^2}{4\pi\epsilon_0}\underbrace{\int \int d\mathbf{r}d\mathbf{r'} \frac{|\phi_i(\mathbf{r})|^2|\phi_j(\mathbf{r'})|^2}{|\mathbf{r} - \mathbf{r'}|}}_{V_{ij,ij}} - \nonumber \\
    & \frac{q^2}{4\pi\epsilon_0}\underbrace{\int \int d\mathbf{r}d\mathbf{r'} \frac{\phi_i(\mathbf{r})^{*}\phi_j(\mathbf{r})\phi_j(\mathbf{r'})^{*}\phi_i(\mathbf{r'})}{|\mathbf{r} - \mathbf{r'}|}}_{V_{ij,ji}}.
\end{align}
Practically, the $\widetilde{V}_{ij}$ terms in Eq.~(\ref{eq:Coulomb_matrix}) are evaluated by solving Poisson's equation and determining the electrostatic potential $U_i(r)$ induced by a pseudo charge centered at $\mathbf{R}_i$ and having a spatial distribution equal to the square of the $\phi_i(\mathbf{r'})$ wavefunction, i.e., $|\phi_i  (\mathbf{r'})|^2$. In this approach, $U_i(r)$ is defined as 
\begin{equation}\label{eq:coulomb_poisson}
 U_i(r) = \frac{q}{4\pi\epsilon_0}\int d\mathbf{r'} \frac{|\phi_i  (\mathbf{r'})|^2}{|\mathbf{r-r'}|}, 
\end{equation}
which is equivalent to the solution of Poisson's equation. Hence, to tackle Eq.~(\ref{eq:coulomb_poisson}) we take advantage of the Poisson equation solver from the GPAW tool \cite{Mortensen2024} and repeat this operation for all indices $i$. Finally, $V_{ij,ij}$ is computed by performing the following volume integration
\begin{equation}\label{eq:coulomb_integral}
\widetilde{V}_{ij,ij} = q\int d\mathbf{r}  U_i(r)  | \phi_j  (\mathbf{r})|^2.
\end{equation}
In this work, we further neglect long-range elements of the two-index interaction matrix $\mathrm{\mathbf{\widetilde{V}}}$. Indices $i$ and $j$ refer to the orbitals of atoms situated at positions $\mathbf{R}_i=(x_i,y_i,z_i)$ and $\mathbf{R}_j=(x_j,y_j,z_j)$, respectively. Only interaction distances satisfying $|_i-_j|\leq r_{max}$ are kept, which ensures that $\mathbf{\widetilde{V}}$ also possesses a block-tridiagonal sparsity pattern similar to that of $\mathbf{H}$. The off-diagonal blocks of $\mathbf{\widetilde{V}}$ contain interactions between adjacent device cells, i.e., $i$ and $j$ belong to two different cells. Thanks to the periodicity of MLWFs, they can be computed by shifting either $\phi_i$ or $\phi_j$ to a neighboring cell before computing the integral in Eq.~(\ref{eq:coulomb_integral}). As in the Hamiltonian case, the knowledge of a set of diagonal and off-diagonal blocks is sufficient to scale  $\mathbf{\widetilde{V}}$ up to the device dimensions.

To compensate for the reduction of $\mathrm{\mathbf{\widetilde{V}}}$ to a two-index tensor and for the omission of long-range iterations, we assume that our bare Coulomb matrix is embedded in a dielectric environment. It is, therefore, scaled by a factor $\epsilon_R$, i.e., $\mathrm{\mathbf{\widetilde{V}}}\rightarrow\mathrm{\mathbf{\widetilde{V}}}/\epsilon_R$. This factor produces the same effect as external dielectric screening coming, for example, from an underlying substrate or a surrounding oxide. Its value can be calibrated by matching the DFT+NEGF+scGW bandgap of a given material to the one resulting from a reference G$_0$W$_0$ calculation or from experiments.

\subsection{Self-Consistent DFT+NEGF+GW Equations \label{subsec:SC_equationschapter}}
To perform a quantum transport simulation within the NEGF formalism, the following equations must be solved for the energy-dependent retarded ($\mathbf{G}^R(E)$), lesser ($\mathbf{G}^<(E)$), and greater ($\mathbf{G}^>(E)$) Green's functions \cite{negf}:
\begin{equation}
\left(E-\mathbf{H}-\mathbf{V_{ext}}-\mathbf{\Sigma}^{R}_{B}(E)- \mathbf{\Sigma}^{R}_{GW}(E)\right) \cdot \mathbf{G}^{R}(E)  = \mathbf{I},
\label{eq:GR}
\end{equation}
\begin{equation}
\label{eq:lessgtrGLGtilda}
   \mathbf{G}^{\lessgtr}(E) = \mathbf{G}^R(E) \lbrack \mathbf{\Sigma}^{\lessgtr}_{B}(E) + \mathbf{\Sigma}^{\lessgtr}_{GW}(E) \rbrack \cdot {(\mathbf{G}^{R})^{\dagger}}(E).
\end{equation}
In Eqs.~(\ref{eq:GR}) and (\ref{eq:lessgtrGLGtilda}), all matrices have the same size as $\mathbf{H}$ ($N_{AO}\times N_{AO}$), $E$ is the electron energy, $\mathbf{I}$ the identity matrix, while the $\Sigma$'s represent self-energies that can be of boundary ($B$) or scattering ($GW$) type. The diagonal matrix $\mathbf{V_{ext}}$ contains the electrostatic potential energy induced by external sources, e.g., applied voltages or doping. It can be calculated via Poisson's equation
\begin{equation}\label{eq:Poisson}
    \nabla \left(\mathbf{\epsilon} \cdot \nabla U_{ext}  \right) = -q(p - n - N_A^{-} + N_D^{+})
\end{equation}
with appropriate boundary conditions and $-qU_{ext}$ as the diagonal entries of $\mathbf{V_{ext}}$. In Eq.~(\ref{eq:Poisson}), $\epsilon$ is the spatially-dependent dielectric permittivity of the structure under consideration, $N_A^{-}$ and $N_D^{+}$ are the ionized acceptor and donor concentrations, respectively, which remain constant throughout the simulation process, while $p$ and $n$ are the hole and electron densities. They can be derived alongside the density-of-states (DOS) from the lesser/greater Green's functions
\begin{align}
    \label{eq:DOS}
    DOS(E,\mathbf{R}_{\xi}) &= - \frac{1}{\pi}\mathfrak{Im} \sum_{n} G_{nn}^{R}(E),\\
    \label{eq:eleDensity}
    n(\mathbf{R}_{\xi}) &= -2i \sum_{n}\int_{CB} \frac{dE}{2\pi}G_{nn}^{<}(E), \\
    \label{eq:holeDensity}
    p(\mathbf{R}_{\xi}) &= 2i \sum_{n}\int_{VB} \frac{dE}{2\pi}G_{nn}^{>}(E),\\
\end{align}
The sums in Eqs.~\eqref{eq:eleDensity}-\eqref{eq:holeDensity} run over the indices referring to all Wannier orbitals associated with an atom located at $\mathbf{R}_{\xi}$. The factor 2 accounts for the spin degeneracy. Due to the high localization of the MLWFs, the electrons and holes are approximated as point charges with a $\delta$-like spatial distribution. Since the energy integration for $n(\mathbf{R}_{\xi})$ and $p(\mathbf{R}_{\xi})$ is performed over the conduction (CB) and valence band (VB) states, respectively, the obtained electron and hole densities can be seen as the excess charges of the system. We further assume that the influence of the VB electron population and of the positive charges coming from the atomic cores is already accounted for in the MLWF Hamiltonian and is thus no longer considered in Poisson's equation. The validity of such a scheme has been discussed in \cite{Duflou2023}.

As can be seen in Eqs.~(\ref{eq:GR}) and (\ref{eq:lessgtrGLGtilda}) the Green's functions $\mathbf{G}(E)$ depend on the GW self-energies, $\mathrm{\mathbf{\Sigma}}_{GW}(E)$, which are used to model e-e interactions. They are also referred to as the exchange-correlation (XC) self-energies and include the Fock diagram (exchange) and a first-order diagram accounting for non-local in-time effects (correlation) \cite{Stefanucci}. The lesser/greater SSEs are given by \cite{PhysRevB.77.115333} 
\begin{equation}
\label{eq:siglg}
\Sigma^{\lessgtr}_{GW,ij}(E) = i \int dE' \lbrack G^{\lessgtr}_{ij}(E')W^{\lessgtr}_{ij}(E-E') \rbrack,
\end{equation}
where $\mathbf{W}^{\lessgtr}(E')$ is the dynamically screened Coulomb interaction. The indices $i$ and $j$ refer to a pair of Wannier orbitals. The retarded GW self-energy, $\mathbf{\Sigma}^{R}_{GW}(E)$, consists of the sum of an exchange
\begin{equation}
\label{eq:sigr_x_ed}
\Sigma^{R}_{x, ij} = i\int_{E}dE \text{} G^{<}_{ij}(E)\widetilde{V}_{ij}
\end{equation}
and correlation
\begin{equation}
\label{eq:PI}
\Sigma^{R}_{corr, ij}(E) = - \frac{i}{2}\Gamma_{ij}(E) + \mathcal{P} \text{ } \int_{E} \frac{dE'}{2\pi}\frac{\Gamma_{ij}(E')}{E-E'}
\end{equation}
term. In Eq.~(\ref{eq:PI}), $\mathcal{P}$ denotes the principal part of the integral and $\Gamma_{ij}$ is the broadening function defined as 
\begin{equation}
    \Gamma_{ij} = i \lbrack \Sigma^{>}_{GW,ij} - \Sigma^{<}_{GW,ij} \rbrack.
\end{equation}
Details about the derivation of these equations are provided in Appendix~\ref{app:gw}.

To compute the GW self-energies, the screened Coulomb interactions $\mathbf{W}$ are required. The retarded component of this quantity obeys a Dyson equation with a similar form as Eq.~(\ref{eq:GR})
\begin{equation}
\label{eq:WR}
\underbrace{\left( \mathbf{I} - \underbrace{\mathbf{\widetilde{V}}\cdot\mathbf{P}^R(E)}_{\mathbf{K}^{R}(E)} -\mathbf{S}^{R}_{B}(E) \right)}_{:=\mathbf{M}^{R} = (\mathbf{X}^R (E))^{-1}}\cdot\mathbf{W}^R (E)=\mathbf{\widetilde{V}}.
\end{equation}
The lesser and greater components are obtained with
\begin{align}
\label{eq:WLG}
\mathbf{W}^{\lessgtr}(E) &= \mathbf{X}^R(E)\lbrack \underbrace{\mathbf{\widetilde{V}} \cdot\mathbf{P}^{\lessgtr}(E)\cdot\mathbf{\widetilde{V}^{T}}}_{:=\mathbf{L}^{\lessgtr}(E)} + \nonumber \\
&\mathbf{S}^{\lessgtr}_{B}(E)\rbrack\cdot{(\mathbf{X}^{R})^{\dagger}}(E).
\end{align}
In both equations, a new physical quantity is introduced, the polarization $\mathbf{P}$ in its retarded, lesser, and greater form. We have that \cite{PhysRevB.77.115333}
\begin{equation}
\label{eq:Plessgtr}
P^{\lessgtr}_{ij} = - i \int dE' \lbrack G^{\lessgtr}_{ij}(E')G^{\gtrless}_{ji}(E'-E) \rbrack.
\end{equation}
Instead of using the exact expression for the retarded component of $\mathbf{P}$, we approximate it as \cite{lake}
\begin{equation}
\label{eq:Pr}
\mathbf{P^{R}} \approx \frac{1}{2} \left( \mathbf{P}^{>} - \mathbf{P}^{<} \right), 
\end{equation}
which has been found to be numerically more stable, without significantly affecting the physical results.

Going back to Eqs.~(\ref{eq:WR}) and (\ref{eq:WLG}), the $\mathbf{S}^{R,\lessgtr}_{B}$ represent the boundary ``self-energies'' of $\mathbf{W}$. Their derivation and importance are discussed in Section \ref{subsec:boundaries}. Finally, note that the definition of $\mathbf{X}^R(E)$ in Eq.~\eqref{eq:WR} serves two purposes. Firstly, it provides a direct analogy to $\mathbf{G}^R$ with its $\mathbf{M^R}\cdot\mathbf{X}^R(E)=\mathbf{I}$ form. Secondly, it facilitates the computation of $\mathbf{W}^{\lessgtr}(E)$ in Eq.~(\ref{eq:WLG}). 

In this work, the systems of linear equations in (\ref{eq:GR}), (\ref{eq:lessgtrGLGtilda}), (\ref{eq:WR}), and (\ref{eq:WLG}) are solved with the so-called recursive Green's function (RGF) algorithm \cite{rgf}. RGF had to be adapted to handle the computation of the $\mathbf{W}$ components, as explained in Appendix \ref{app:RGF_new}. At the same time, the convolutions in Eqs.~(\ref{eq:siglg}) and (\ref{eq:Plessgtr}) are not calculated in the energy domain, but in the time domain after a fast Fourier transform (FFT) \cite{PhysRevB.77.115333}. The results are obtained after applying an inverse FFT. The principal part of the integral in Eq.~(\ref{eq:PI}) is also taken care of in the time domain after an FFT. All these numerical operations have been massively parallelized and ported to graphics processing units (GPUs) to accelerate their execution. The implementation strategy is outlined in \cite{10.1109/SC41406.2024.00069}.

Overall, Eqs.~\eqref{eq:GR}-\eqref{eq:Pr} constitute the full set of equations of our DFT+NEGF+scGW scheme. They must be solved self-consistently in a loop until convergence is reached, i.e., within each iteration of this loop, a $\mathbf{G}\rightarrow\mathbf{P}\rightarrow\mathbf{W}\rightarrow{\Sigma}$ sequence must be executed. This procedure is known as the self-consistent Born approximation (SCBA) \cite{Kadanoff2018}. Once, for example, the changes in the $\mathbf{G}$ matrices between two consecutive iterations do not exceed a pre-defined criterion any more, physical observables such as the DOS, electron and hole densities can be evaluated with Eqs.~\eqref{eq:DOS} to \eqref{eq:holeDensity}. 

The electronic and energy current densities are other relevant quantities. Typically, they are computed with the Meir-Wingreen formula, which gives the current flowing from an electrode $\alpha$ inside the device of interest \cite{PhysRevLett.68.2512}. Hence, if there are two contacts, two values can be produced. Ideally, the current should be evaluated at each unit cell constituting the device structure to gain insight into both its spatial and spectral resolution. For that purpose, we generalized the Meir-Wingreen equation
\begin{align}
\label{eq:meir-wingreen}
        I_{n} = -\frac{2q}{\hbar} \mathfrak{Re} \int_{-\infty} ^ {\infty}  \frac{dE}{2\pi} \text{ } \mathrm{tr}& \{ \mathbf{\widetilde{\Sigma}}^{>}_{B,n}(E)\mathbf{G}^{<}_{n,n}(E) - \nonumber \\ 
        &  \mathbf{G}^{>}_{n,n}(E)\mathbf{\widetilde{\Sigma}}^{<}_{B,n}(E) \}.
\end{align}
Here, $I_{n}$ is the electronic current flowing through the $n^{\mathrm{th}}$ unit cell, the pre-factor 2 accounts for the spin degeneracy, the $\mathbf{G}^{\lessgtr}_{n,n}$ are the $n^{\mathrm{th}}$ diagonal blocks of the $\mathbf{G}^{\lessgtr}$ matrices, whereas the $\mathbf{\widetilde{\Sigma}}^{\lessgtr}_{B,n}$ self-energy blocks propagate the influence of the boundary $B\in\{L,R\}$ to the $n^{\mathrm{th}}$ unit cell and thus allow to compute the current directly at this location. The generalization of the Meir-Wingreen formula for $I_{n}$ is derived in Appendix \ref{app:current} where $\mathbf{\widetilde{\Sigma}}^{\lessgtr}_{B,n}$ is defined in Eq.~(\ref{eq:sigmab_mw}). A similar equation can be used for the energy current
\begin{align}
\label{eq:energy_meir-wingreen}
        I_{n} = -\frac{2q}{\hbar} \mathfrak{Re} \int_{-\infty} ^ {\infty}  \frac{dE}{2\pi} \text{ } (E-E_{F}) \text{ } \mathrm{tr}& \{ \mathbf{\widetilde{\Sigma}}^{>}_{B,n}(E)\mathbf{G}^{<}_{n,n}(E) - \nonumber \\ 
        &  \mathbf{G}^{>}_{n,n}(E)\mathbf{\widetilde{\Sigma}}^{<}_{B,n}(E) \},
\end{align}
where $E_{F}$ is the Fermi level of the contact from which electrons are injected into the simulated device. As long as the entries of the screened Coulomb interaction matrix satisfy $W_{ij}^{<}(E)=W_{ij}^{>}(-E)$, both currents are conserved.

\subsection{Open Boundary Conditions \label{subsec:boundaries}}
In quantum transport calculations, semi-infinite leads (contacts) are attached to a central region called ``device'' such that electrons can enter and leave the simulation domain with a probability that depends on their occupation number \cite{datta_1995}. The resulting open boundary conditions (OBCs) are cast into the $\mathbf{\Sigma}^{R}_{B}(E)$ and $\mathbf{\Sigma}^{\lessgtr}_{B}(E)$ self-energies in Eqs.~(\ref{eq:GR}) and (\ref{eq:lessgtrGLGtilda}), respectively. In structures with two contacts, as in this work, only the first and last block of these matrices are different from 0. The retarded component is then defined as \cite{Bowen1995}
\begin{equation}
\label{eq:sigb}
\mathbf{\Sigma}^{R}_{B(i,i)}=\mathbf{A}_{i,j}\left( \mathbf{A}_{j,j} + \mathbf{A}_{i,j}\mathbf{\Phi}_C \mathbf{\Lambda}^{-1}_C \mathbf{\Phi}^{-1}_C  \right) \mathbf{A}_{j,i},
\end{equation}
assuming that (i) the device is made of $N$ unit cells, (ii) each unit cell in the left (right) contact is identical to the first (last) one, (iii) the pair $(i,j)$ is either equal to $(1,0)$=$(2,1)$ or $(N,N+1)$=$(N-1,N)$, (iv) $C$ stands for $L$ (left contact) or $R$ (right contact), (v) plane-waves are injected from the leads into the device, and (vi) $\mathbf{A}_{i,j}$ is the $(i,j)$ block of the matrix $\left(E-\mathbf{H}-\mathbf{V_{ext}}-\mathbf{\Sigma}^{R}_{GW}(E)\right)$ with size $d$ (number of orbitals per contact unit cell). 

Moreover, in Eq.~(\ref{eq:sigb}), the matrix $\mathbf{\Phi}_C=\left( \phi_{1}, \hdots , \phi_{c} \right)$ of size $d\times c$ contains the $c\ll d$ modes $\phi_{i}$ of size $d$ that propagate or exponentially decay in contact $C$, away from the device (reflected states). The diagonal matrix $\Lambda_C$ is made of all phase factors $e ^ {ik_i}$ with wave vector $k_i$ corresponding to the $\phi_{i}$ modes. The calculation of the $k_i$ and $\phi_{i}$ takes the form of a non-linear eigenvalue problem (NLEVP) \cite{brueck}. Since only the phase factors satisfying $1/R\leq\left|\mathrm{exp}(ik))\right|\leq R$ with 100$\leq R \leq$1000 are required, contour integral techniques such as the Beyn algorithm \cite{beyn} can be recalled to efficiently compute them. 

The challenge comes from the fact that $\mathbf{\Phi}^{-1}_C$ (size: $c\times d$) is needed in Eq.~(\ref{eq:sigb}) and since $\mathbf{\Phi}_C$ is a rectangular matrix, its inverse cannot be directly obtained. It can, however, be determined from
\begin{equation}
\mathbf{\Phi}_C^{-1} = \left(\mathbf{\Phi}^{\dagger}_C\cdot\mathbf{\Phi}_C \right)^{-1}\cdot \mathbf{\Phi}^{\dagger}_C.
\end{equation}
The lesser/greater boundary self-energies can be constructed from the retarded one by applying the fluctuation-dissipation theorem \cite{2013stefanucci_chap09}. First, the contact-broadening function
\begin{equation}
\mathbf{\Gamma}_{i,i}(E) = i\left( \mathbf{\Sigma}^{R}_{B(i,i)}(E) - {(\mathbf{\Sigma}^{R}_{B(i,i)})^{\dagger}(E)} \right),
\end{equation}
is defined from which
\begin{align}\label{eq:sigmaLB}
\mathbf{\Sigma}^{<}_{B(i,i)}(E) &= i \mathbf{\Gamma}_{i,i}(E)f(E, E_{FC}), \\
\label{eq:sigmaGB}
\mathbf{\Sigma}^{>}_{B(i,i)}(E) &= -i \mathbf{\Gamma}_{i,i}(E)(1-f(E, E_{FC})),
\end{align}
can be derived. In Eqs.~\eqref{eq:sigmaLB} and~\eqref{eq:sigmaGB}, $f(E, E_{FC})$ is the Fermi-Dirac distribution at energy $E$ for contact $C$. 

Since the simulated system is also open for the screened Coulomb interaction $\mathbf{W}$, proper OBCs must be introduced for this quantity too. To the best of our knowledge, this is the first time that this is done in quantum transport calculations with e-e interactions. The necessity of including OBCs for $\mathbf{W}$ is demonstrated in Appendix \ref{app:obcw}: Without them, the results are shown to be inaccurate.

The $\mathbf{S}^R_B$ boundary matrix in Eq.~(\ref{eq:WR}) has the same size and structure as $\mathbf{\Sigma}^R_B$ and can thus be computed in a similar way with the help of Eq.~(\ref{eq:sigb}). The only difference comes from the $\mathbf{A}_{i,j}$ blocks, which must be extracted from the $(\mathbf{I}-\mathbf{K}^R(E))$ matrix. The latter is also at the core of the NLEVP that is solved with the Beyn algorithm and that produces $\mathbf{\Phi}_C$.

The situation is more complex for $\mathbf{W}^{\lessgtr}$ and its boundary matrix $\mathbf{S}^{\lessgtr}_B$. Although the fluctuation-dissipation theorem also applies to the screened Coulomb interaction, which is a Boson, we found out that this is not the best approach to define OBCs for $\mathbf{W}^{\lessgtr}$. In Appendix \ref{app:RGF_new} about the extension of the RGF algorithm to $\mathbf{W}^{\lessgtr}$, a relationship between the right-connected $\mathbf{w}^{\lessgtr}_{n+1,n+1}$ and $\mathbf{w}^{\lessgtr}_{n,n}$ matrices is established as part of Eq.~(\ref{eq:wleftconn}). The meaning of a right-connected block $n$ in the context of RGF is that all physical effects occurring between the $n^{\mathrm{th}}$ device unit cell and $+\infty$ are accounted for, but there is no connection with what happens at cells with index $i<n$. 

This property can be leveraged to find an expression for $\mathbf{S}^{\lessgtr}_B$. Since the semi-infinite leads are initially disconnected from the device, the recursion in Eq.~(\ref{eq:wleftconn}) is valid in these regions. Furthermore, as the contacts are typically made of the repetition of the same unit cell, $\mathbf{w}^{\lessgtr}_{n,n}$ and $\mathbf{w}^{\lessgtr}_{n+1,n+1}$ have to be equal. As a consequence, Eq.~(\ref{eq:wleftconn}) forms a closed set of equations in the leads
 
\begin{align}
\label{eq:surface_w}
    \mathbf{w}^{\lessgtr}_{i, i} &= \mathbf{x}^R_{i, i} \lbrack \mathbf{L}^{\lessgtr}_{i, i} + \mathbf{M}_{i, j} \mathbf{w}^{\lessgtr}_{i,i} {\mathbf{M}^{\dagger}_{i, j}} - 
    \left( \mathbf{a}^{\lessgtr}_{i, i} - {\mathbf{a}^{\lessgtr}_{i, i}}^{\dagger} \right)
       \rbrack (\mathbf{x}^R_{i, i}) ^{\dagger},
\end{align}
with the $(i,j)$ pair equal to $(1,0)$ for the left contact and $(N,N+1)$ for the right one. An algorithm to solve this Lyapunov-type equation is proposed in Appendix \ref{app:obcw}. Once $\mathbf{w}^{\lessgtr}_{i, i}$ is available, it serves as input to calculate $\mathbf{S}^{\lessgtr}_B$
\begin{align}
    \mathbf{S}^{\lessgtr}_{B(i, i)} &= \mathbf{M}_{i, j} \mathbf{w}^{\lessgtr}_{i,i} {\mathbf{M}^{\dagger}_{j, i}} - \left( \mathbf{a}^{\lessgtr}_{i, i} - {\mathbf{a}^{\lessgtr}_{i, i}}^{\dagger} \right).
\end{align}
This expression directly follows from Eqs.~(\ref{eq:wleftconn}) and (\ref{eq:surface_w}) after separating the influence of scattering ($\mathbf{L}^{\lessgtr}_{i, i}$) from the rest, which gives rise to $\mathbf{S}^{\lessgtr}_{B(i, i)}$. 

\section{Results}\label{sec:Results}

\subsection{(8,0) single-wall carbon nanotube (SWCNT)}

\begin{figure}[t]
    \centering
        \includegraphics[width=0.45\textwidth]{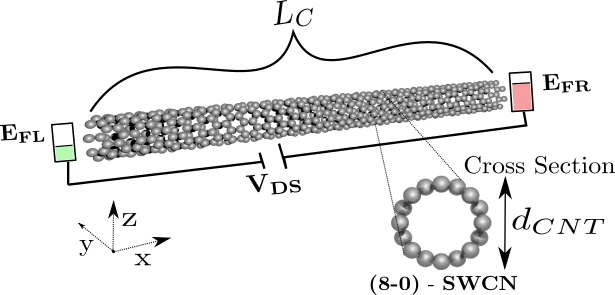}
              \caption{Schematic view of a (8,0)-SWCNT with a length $L_{C}$=20.5 nm, a diameter$d_{CNT}$=0.63 nm, and a total number of atoms $N_A$=1536. A bias $V_{DS}$ is applied between the left- and right extremities of the CNT, which are characterized by a Fermi level $E_{FL}$ and $E_{FR}$, respectively. Transport occurs along the $x$-axis.}
            \label{fig:CNT_sketch}
\end{figure}

The DFT+NEGF+scGW model described in Section~\ref{sec:Method} is now applied to a (8,0) single-wall carbon nanotube (SWCNT), as shown in Fig.~\ref{fig:CNT_sketch}, to assess its validity. In particular, the transport properties of this structure are analyzed as a function of the maximum interaction distance $r_{max}$ between two atoms/orbitals. As mentioned in Section \ref{subsec:CM}, $r_{max}$ leads to a truncation of the bare Coulomb matrix $\mathbf{\widetilde{V}}$ as well of the polarization function $\mathbf{P}$ and scattering self-energies $\mathbf{\Sigma}$. Hence, all these matrices have a block-tridiagonal sparsity pattern. The latter is key to the application of the RGF algorithm.

The SWCNT considered here has a diameter $d_{CNT} = 0.63$ nm and a length $L_C = 20.5$ nm that can be divided into 48 unit cells along the transport axis ($x$), each of them with a width $l_{uc} = 0.43$ nm. To examine the convergence of our quantum transport scheme we assume either flat-band conditions (all entries of $\mathbf{V_{ext}}$ are equal to 0) or a linear potential drop between two flat regions of length $L_{side}$=3.5 nm and separated by 18.5 nm. In this case, a voltage $V_{DS}$ is set, thus splitting the left ($E_{FL}$) and right ($E_{FR}=E_{FL}-qV_{DS}$) Fermi levels. All simulations are run at room temperature. Note that Poisson's equation is not solved in these simulations as our goal is to demonstrate the numerical convergence of our DFT+NEGF+scGW approach, not yet to investigate physical effects. This will be done in Section \ref{sec:sinwfet}.

The bandstructure of the (8,0)-SWCNT is shown in Fig.~\ref{fig:flatband_CNT}(a), as computed with the plane-wave VASP code \cite{vasp} and after transformation into MLWFs \cite{wannier90}. One single $p_z$-like orbital per carbon atom is retained. It is sufficient to obtain an excellent agreement between both sets of bands around the band gap of the CNT. The energy window over which the MLWF bands extends goes from -11.6 eV up to 0.9 eV. To ensure accurate results, in Eq.~(\ref{eq:GR}), we define an energy vector $E$ with values between $E_{min} = -35$ eV and $E_{max} = +25$ eV and a resolution $\Delta_{E} = 5$ meV. Such a resolution is necessary to describe all features of the SWCNT bandstructure, whereas the large range ensures that the widening of the conduction and valence bands is taken into account and does not induce numerical issues when performing the convolutions in Eqs.~\eqref{eq:siglg}, \eqref{eq:PI}, and \eqref{eq:Plessgtr} with FFT. 

\begin{figure}[t]
\centering
        \includegraphics[width=0.45\textwidth]{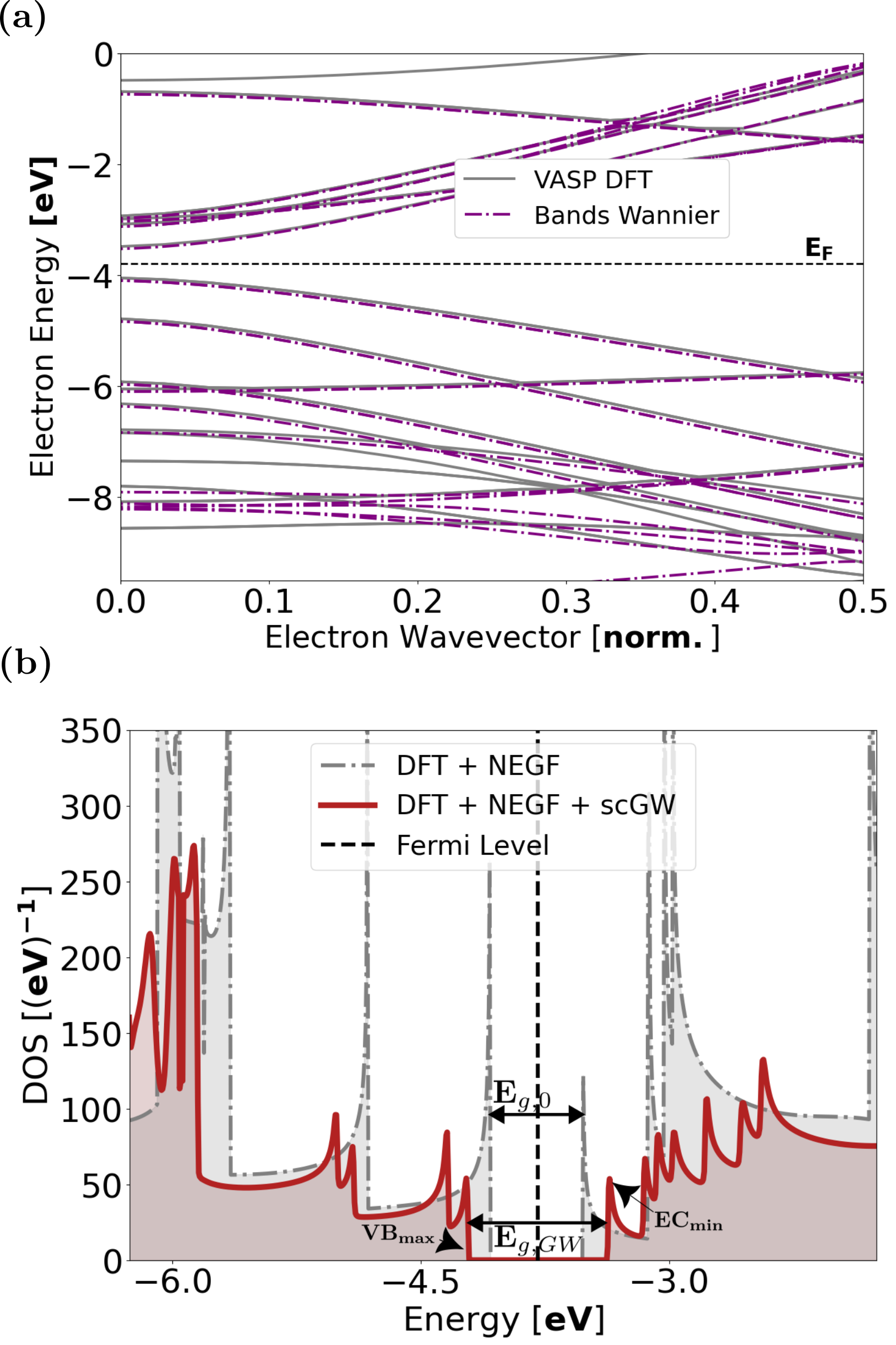}

\caption[Two numerical solutions]{\label{fig:flatband_CNT} \textbf{(a)} Bandstructure of the (8,0)-SWCNT as computed with VASP (solid blue lines) and the MLWF basis (dashed red line). \textbf{(b)} Local DOS of the (8,0)-SWCNT under flat-band conditions. The gray shaded area represents the case of DFT+NEGF, the red one DFT+NEGF+scGW. The Fermi level, valence band maximum ($VB_{max}$), conduction band minimum ($CB_{min}$), pure DFT band gap ($E_{g,0}$), and band gap with GW ($E_{g,GW}$) are indicated in the plot.}
\label{fig:flatband_CNT}
\end{figure}

First, the SWCNT sample is analyzed without any applied bias, under flat-band conditions. The Fermi levels $E_{FL}$ and $E_{FR}$ are fixed in the middle of the bandgap. After running our DFT+NEGF+scGW algorithm till convergence, the computed DOS per unit cell is extracted with Eq.~(\ref{eq:DOS}) and compared to its pure DFT+NEGF counterpart in Fig.~\ref{fig:flatband_CNT}(b). A bandgap increase from $E_{g, 0} = 0.56$ eV to $E_{g, GW} = 0.96$ eV is observed, when electron-electron interactions are included. As explained in Section~\ref{subsec:CM}, the parameter $\epsilon_R$ with which the bare Coulomb matrix is scaled can be used to compensate for the missing entries of $\mathbf{\widetilde{V}}$ and to model the influence of external dielectric screening. The dielectric environment can indeed significantly affect the bandgap of CNTs \cite{Aspitarte2017}. In fact, GW calculations of (8,0)-SWCNTs in the presence of a hexagonal boron nitride (h-BN) substrate have shown that, depending on h-BN/CNT separation distance, the bandgap varies between 0.75 and 1.45 eV \cite{Lanzillo2014}. We therefore set $\epsilon_R$=3.0, in combination with $r_{max}$=0.86 nm, which together yields a band gap in this range.

\begin{figure*}[!t]
\centering
        \includegraphics[width=1.0\textwidth]{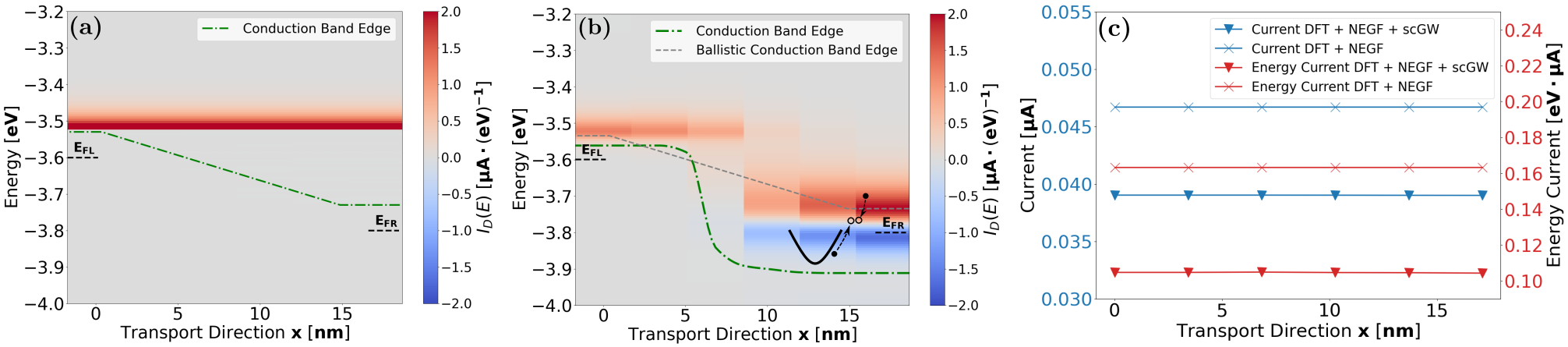}

\caption[Two numerical solutions]{\label{fig:spectral_currents_CNT} Position- and energy-resolved electronic current when a bias of $V_{DS} = 0.2$ V is applied to the SWCNT structure depicted in Fig.~\ref{fig:CNT_sketch}. The position of the Fermi levels is indicated by dashed lines, the approximate conduction band edge with green dash-dots. Red-shaded areas represent electrons flowing from left to right, and blue-shaded areas the opposite. \textbf{(a)} Ballistic transport with DFT+NEGF. \textbf{(b)} Transport with electron-electron interactions through DFT+NEGF+scGW. The sketch on the right represents energy-conserving e-e scattering processes where filled (empty) circles refer to electrons (empty states). The black dashed line represents the applied linear potential drop through $\mathbf{V_{ext}}$ in Eq.~(\ref{eq:GR}). \textbf{(c)} Position-dependent electronic (blue) and energy (red) currents corresponding to the simulations in sub-plots \textbf{(a)} and \textbf{(b)}.}
\label{fig:spec_current_CNT_GW}
\end{figure*}

Next, the SWCNT is driven out of equilibrium by applying an external bias $V_{DS}$=0.2 V and a linear potential drop. The Fermi levels $E_{FL}$ and $E_{FR}$ are fixed 75 meV below the conduction band edge to emulate n-type doping. The corresponding position- and energy-resolved current density is reported in Fig.~\ref{fig:spec_current_CNT_GW}(a) in case of pure ballistic DFT+NEGF simulations and in Fig.~\ref{fig:spec_current_CNT_GW}(b), after convergence of the $\mathbf{G}\rightarrow\mathbf{P}\rightarrow\mathbf{W}\rightarrow{\mathbf{\Sigma}}$ self-consistent cycle, i.e., in the presence of e-e scattering. The imposed electrostatic potential energy is indicated in both sub-plots to help interpret these results. It should be emphasized that electronic and energy current conservation is ensured in all types of calculations (ballistic and with e-e scattering), as demonstrated in Fig.~\ref{fig:spec_current_CNT_GW}(c). 

In the ballistic limit, carriers are injected from the left contact and propagate coherently above the conduction band edge to the right. Since the external electrostatic potential is kept constant, the band edge remains perfectly linear. When e-e interactions are turned on, the current distribution exhibits an entirely different picture. First, it can be seen that the conduction band (CB) edge, extracted from the position-dependent DOS, changes significantly, as compared to ballistic transport, and does not coincide with the applied linear potential drop any more. The CB modulation can be attributed to local variations of the charge density, the screening effects they induce, and the resulting band gap changes that are captured by the real part of $\Sigma^{R}_{GW}$. Because Poisson's equation is not solved in this simulation, the carrier density becomes higher on the right side of the SWCNT, which is a consequence of the imposed electrostatic potential. 

More interestingly, electron-electron scattering leads to inelastic transport phenomena.
As indicated in the sketch of Fig.~\ref{fig:spec_current_CNT_GW}(b), incoming electrons from the left contact (black circles) with energies around $E_{FL}$ become ``hot'' upon arrival on the right side of the SWCNT. There, they scatter with electrons located below $E_{FR}$ and transfer energy to them so that they can move above $E_{FR}$. The empty states left behind are refilled by the right contact, which is at thermal equilibrium, resulting in a negative current represented by a blue-shaded area. This notable effect of e-e scattering is expected to be more pronounced in highly degenerate 1D structures such as SWCNTs due to their enhanced Coulomb interactions~\cite{thz_excitonics_cnt}.

All out-of-equilibrium results so far were obtained with $\epsilon_R = 1.0$ and $r_{max}$=1.74 nm. The influence of $r_{max}$ is now investigated, while the value of $\epsilon_R$ is kept at 1.0 to maximize the strength of the e-e interactions. As the elements in the 2-index Coulomb matrix $\widetilde{\mathbf{V}}$ slowly decay ($\widetilde{V}_{ij}\propto 1/|\mathbf{R}_i-\mathbf{R}_j|$), it is expected that a high number of off-diagonal elements should be included to ensure that the simulation results do not depend on the choice of the $r_{max}$ value any more. In our implementation, $r_{max}$ determines how many off-diagonal elements are retained in the $\widetilde{\mathbf{V}}$, $\mathbf{P}$, $\mathbf{W}$, and $\mathbf{\Sigma_{GW}}$ matrices and thus sets the band width of these block-tridiagonal matrices.

\begin{figure*}[ht]
\centering
\includegraphics[width=\linewidth]{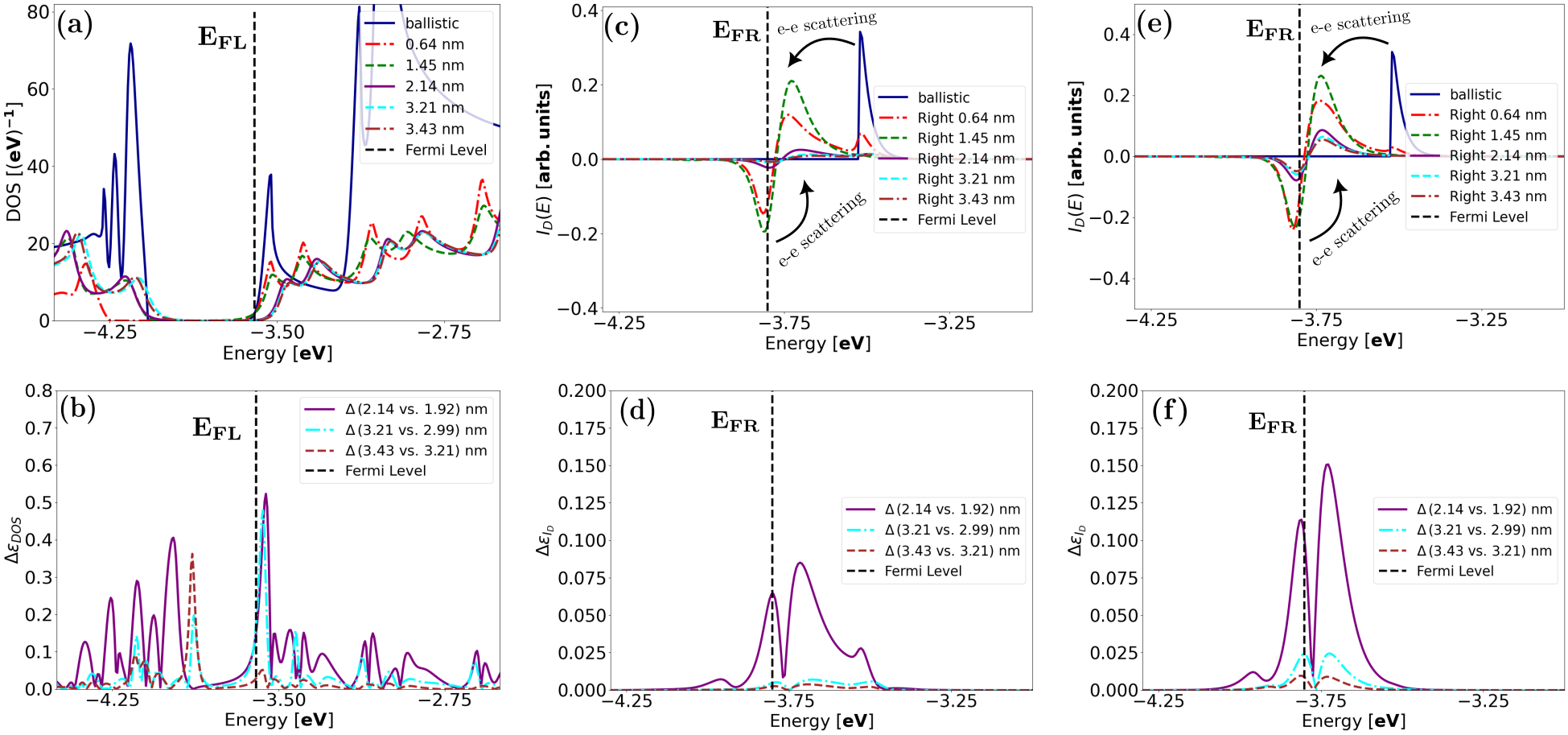}
\caption[Two numerical solutions]{\label{fig:CNT_neighborstudy} Evolution of key physical observables for the (8,0) SWCNT from Fig.~\ref{fig:CNT_sketch} as a function of the $r_{max}$ value applied to the $\widetilde{\mathbf{V}}$, $\mathbf{P}$, $\mathbf{W}$, and $\mathbf{\Sigma_{GW}}$ matrices in our DFT+NEGF+scGW scheme. \textbf{(a)} Energy-resolved local DOS extracted from a unit cell located at $x$=10.25 nm (middle of the SWCNT). \textbf{(b)} DOS relative error $\Delta\epsilon_{DOS}=|DOS(r_{max,1})-DOS(r_{max,2})|/|DOS(r_{max,2})|$ taken between two consecutive values of $r_{max}$: 2.14 and 1.92 nm, 3.21 and 2.99 nm, and 3.43 and 3.21 nm. \textbf{(c)} Energy-resolved electronic current $I_D$ calculated at $x$=12.5 nm. \textbf{(d)} $I_D$ relative error $\Delta\epsilon_{I_D}=|I_D(r_{max,1})-I_D(r_{max,2})|/|I_D(r_{max,2})|$ extracted under the same conditions as in \textbf{(b)}. \textbf{(e)} Same as \textbf{(c)}, but for the current $I_D$ at $x$=25 nm. \textbf{(f)} Same as \textbf{(d)}, but for the data in \textbf{(e)}.}
\label{fig:CNT_neigh}
\end{figure*}

The DOS and electronic current of the same SWCNT structure as before have been simulated as a function of $r_{max}$ ranging from 0.64 to 3.43 nm. The results are shown in Fig.~\ref{fig:CNT_neigh}, both as absolute numbers and as relative errors $\Delta\epsilon$ between two consecutive $r_{max}$ values. As compared to the ballistic case, the DOS at $x$=12.5 nm significantly broadens when e-e scattering is accounted for (sub-plot (a)). Also, notably, the valence band edge keeps shifting until $r_{max} = 3.21$ nm, while the conduction band edge and its overall shape already stabilize at around $r_{max}$=2.14 nm. To further inspect the evolution of the DOS with respect to $r_{max}$, sub-plot (b) reports $\Delta\epsilon_{DOS}=|DOS(r_{max,1})-DOS(r_{max,2})|/|DOS(r_{max,2})|$ between 2.14 and 1.92 nm, 3.21 and 2.99 nm, as well as 3.43 and 3.21 nm. Clearly, the relative errors decrease as $r_{max}$ increases, although this improvement is more marked in the conduction than in the valence band. We would like to underline that the large relative errors around the valence band edge correspond to regions where the DOS is very small. The absolute errors there are limited. 

The same analysis is conducted for the electronic current, extracted in the middle of the SWCNT ($x$=10.25 nm, Fig.~\ref{fig:CNT_neigh}(c) and (d)) and on its right side ($x$=25 nm, Fig.~\ref{fig:CNT_neigh}(e) and (f)). As soon as e-e interactions are present, the maximum of the current gets closer to the source Fermi level $E_{FL}$ and a second, negative, peak appears. It corresponds to the blue-shaded area in Fig.~\ref{fig:spec_current_CNT_GW}(b) and is more important on the right side of the SWCNT than in the middle. Similarly to the DOS, large variations in the current magnitude and distribution can be observed until $r_{max} = 3.21$ nm. The relative error then rapidly decreases, as depicted in Fig.~\ref{fig:CNT_neigh}(d) and (f).

This convergence analysis clearly indicates that a large interaction distance $r_{max}$ is required to converge the physical observables within less than 5-10 \%. For realistic device simulations with larger and more complex structures than SWCNTs, a $r_{max}$ value greater than 3 nm is currently not possible because of the incurred computational burden and memory consumption, which both explode on today's machines, even on large machines such as Alps as the Swiss National Supercomputing Centre~\cite{Alps}. However, it can be observed in Fig.~\ref{fig:CNT_neigh} that the DOS and current results remain qualitatively very similar, regardless of the choice of $r_{max}$.

\begin{figure}[!t]
    \centering
        \includegraphics[width=0.45\textwidth]{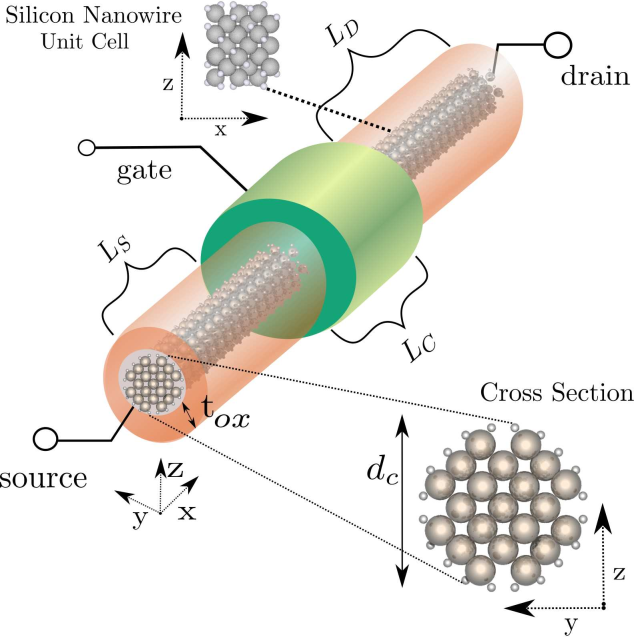}
              \caption{Layout of the gate-all-around silicon nanowire FET simulated in this work. The source, channel, and drain regions are of length $L_S$=14.5 nm, $L_C$=10 nm, and $L_D$=14.5 nm, respectively, while the diameter $d_c$ measures 1 nm. The structure is made of 2952 Si and H atoms, which passivate the surface of the nanowire. Transport occurs along the $x$=$<$100$>$ crystal axis, whereas $y$ and $z$ are directions of confinement. The source and drain extension are doped with an homogeneous donor concentration of 2.5e20 cm$^{-3}$. The nanowire is surrounded by an oxide layer of thickness $t_{ox}$=3 nm and with a dielectric constant $\epsilon_R$=20 around the gate and $\epsilon_R$=3.9 around the source and drain regions.}
            \label{fig:nw_sketch}
\end{figure}

\subsection{Si gate-all-around nanowire FET}\label{sec:sinwfet}
As second application, the GAA Si NWFET displayed in Fig.~\ref{fig:nw_sketch} is investigated with the developed DFT+NEGF+scGW method. The goal is to highlight the influence of electron-electron interactions on the functionality of this type of ultra-scaled nanoelectronic device. The nanowire diameter is set to $d_c$=1 nm, the surface Si atoms are passivated with hydrogen, and the semiconductor channel is surrounded by an oxide layer of thickness $t_{ox}$=3 nm. The source, drain, and gate lengths measure $L_S$=14.5 nm, $L_D$=14.5 nm, and $L_C$=10 nm, respectively. Hence, the structure is made of 72 identical unit cells of width $l_{uc} = 0.547$ nm for a total of $N_A$=2,952 atoms and $N_{AO}$=7,488 MLWF orbitals (4 $sp_3$-like orbitals per Si atom, and 1 $s$-like orbital per H, 104 MLWFs per unit cell). In all calculations, the bare Coulomb matrix $\mathbf{\widetilde{V}}$ is scaled with $\epsilon_R$=5.0 and its interaction range does not exceed $r_{max}$=0.55 nm. These values were chosen so that the simulations remain computationally feasible and the bandgap lies within the desired range (see below). The electron energy vector extends from $E_{min} = -40 \text{ } \mathrm{eV}$ up to $E_{max} = 35 \text{ } \mathrm{eV}$ with a resolution $\Delta_{E} = 5 \text{ } \mathrm{meV}$. Room temperature is assumed in all cases.

\begin{figure}[h]
\centering
        \includegraphics[width=0.45\textwidth]{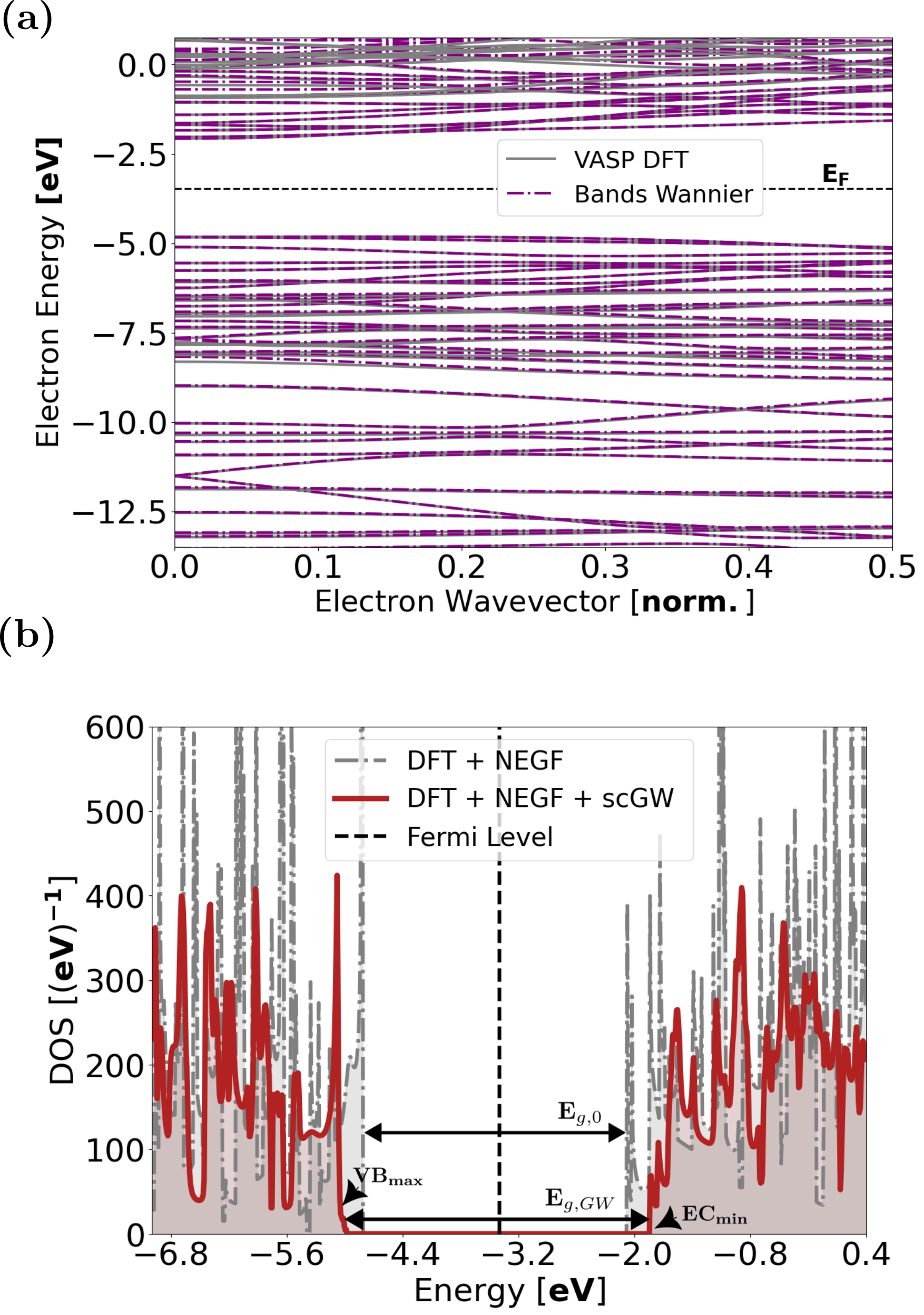}

\caption[Two numerical solutions]{\textbf{(a)} Bandstructure of the Si nanowire from Fig.~\ref{fig:nw_sketch} as computed with the VASP DFT code (solid blue line) and after a transformation into MLWFs (dashed red lines). \textbf{(b)} Local DOS of the same nanowire as before under flat-band conditions. The results obtained with DFT+NEGF (gray) and DFT+NEGF+scGW (red) are compared to each other, highlighting their respective $E_G$ band gap. The vertical dashed line indicates the position of the Fermi level.}
\label{fig:DOS_BS_sinw}
\end{figure}

First, the bandstructure and local DOS of the Si nanowire are examined in detail. Figure \ref{fig:DOS_BS_sinw}(a) shows that an excellent agreement between the DFT and MLWF bands is achieved well above and below the Fermi level, as in the SWCNT case. The DOS resulting from DFT+NEGF and DFT+NEGF+scGW is reported in Fig.~\ref{fig:DOS_BS_sinw}(b), assuming flat-band conditions and an equilibrium Fermi level placed in the middle of the band gap. The latter increases from $E_G$=2.7 to 3.3 eV, when electron-electron scattering is present. Such behavior is expected and the obtained $E_G$ value agrees well with previous theoretical studies \cite{Nolan2006}. Also, the sharp peaks present in the ballistic case broaden, whereas the conduction and valence band edges are smoothed, giving rise to so-called band tails: Electrons and holes can occupy states located within the band gap. 

\begin{figure}[h]
    \centering
        \includegraphics[width=0.45\textwidth]{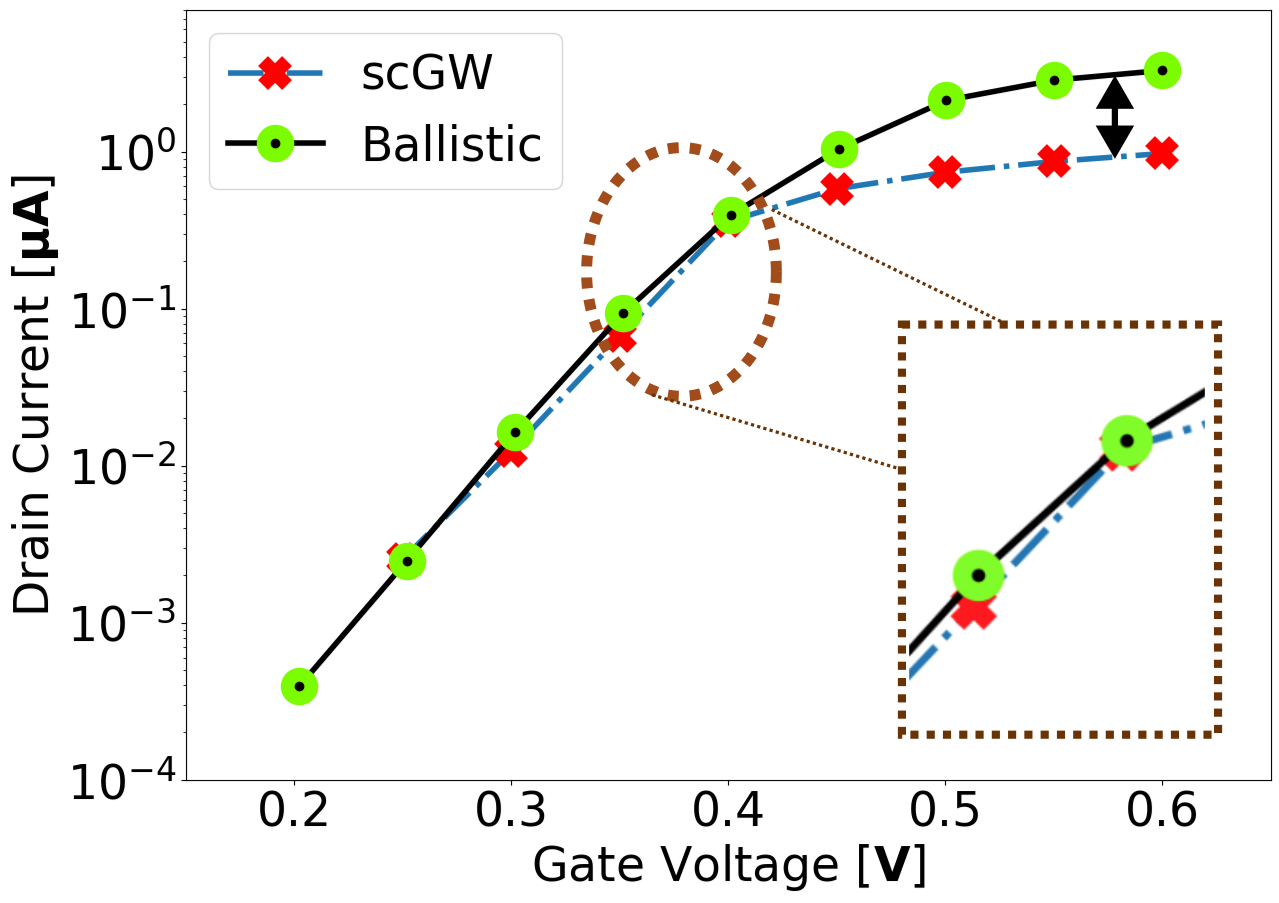}
              \caption{Transfer characteristics $I_{D}$-$V_{GS}$ of the Si NWFET from Fig.~\ref{fig:nw_sketch} at a source-to-drain voltage $V_{DS} = 0.6$ V. The electrical current in the quasi-ballistic limit of transport (black solid line with circles) and in the presence of electron-electron interactions treated in the GW approximation (dashed blue line with crosses) are reported. The double-arrow indicates the ON-state current reduction. The inset magnifies the region around the threshold of the transistor.}
            \label{fig:Id-Vg-curve}
\end{figure}

As next step, the non-equilibrium properties of the Si NWFET are investigated. Its source and drain extensions are doped with a homogeneous donor concentration $N_D = 2.5 \times 10^{20} \text{ } \mathrm{cm ^{-3}}$, its drain-to-source voltage $V_{DS}$ is fixed to 0.6 V, and the gate-to-source voltage $V_{GS}$ is ramped between 0.25 and 0.6 V. The resulting transfer characteristics, computed with DFT+NEGF (quasi-ballistic limit of transport) and DFT+NEGF+scGW (with electron-electron interactions) are displayed in Fig.~\ref{fig:Id-Vg-curve}. To allow for a better comparison, both curves were shifted so that the currents at $V_{GS}$=0.25 V are aligned with each other. Furthermore, to ensure that the results converge at high $V_{GS}$, in the quasi-ballistic case, weak electron-phonon interactions were introduced through the phenomenological model of \cite{Klinkert2020}. It consists of diagonal scattering self-energies of the form
\begin{equation}
\Sigma^{\gtrless}_{ii}(E)=D^2_{ph}\left(n_{\omega}G^{\gtrless}_{ii}(E+\hbar\omega)+(n_{\omega}+1)G^{\gtrless}_{ii}(E-\hbar\omega)\right),
\label{eq:el-phon}
\end{equation}
where $n_{\omega}$ is the Bose-Einstein distribution function and the indices ($i$,$i$) refer to the (row,column) entries of the Green's function and scattering self-energy matrices. We found that a phonon energy $\hbar\omega$=40 meV and a scattering strength $D_{ph}$=25 meV minimize the impact of electron-phonon interactions on the current distribution while enabling the NEGF+Poisson self-consistent loop to converge when $V_{GS}\geq$0.5 V.

Both $I_D$-$V_{GS}$ characteristics agree very well in the subthreshold regime ($V_{GS}$ below 0.4 V), but at higher $V_{GS}$, they start diverging. At $V_{GS}$=0.6 V, the current computed with DFT+NEGF+scGW is about 3.5$\times$ smaller than the quasi-ballistic one. This reduction is caused by backscattering. Each time an electron injected from the source side interacts with another one, it has 50\% probability of continuing its route towards the drain and 50\% to be reflected back to its origin. This feature was also observed in a previous work in which e-e scattering was introduced into a Boltzmann Transport Equation (BTE) solver based on the Monte Carlo algorithm~\cite{Fischetti2001}.

\begin{figure*}[ht]
\centering
\includegraphics[width=\linewidth]{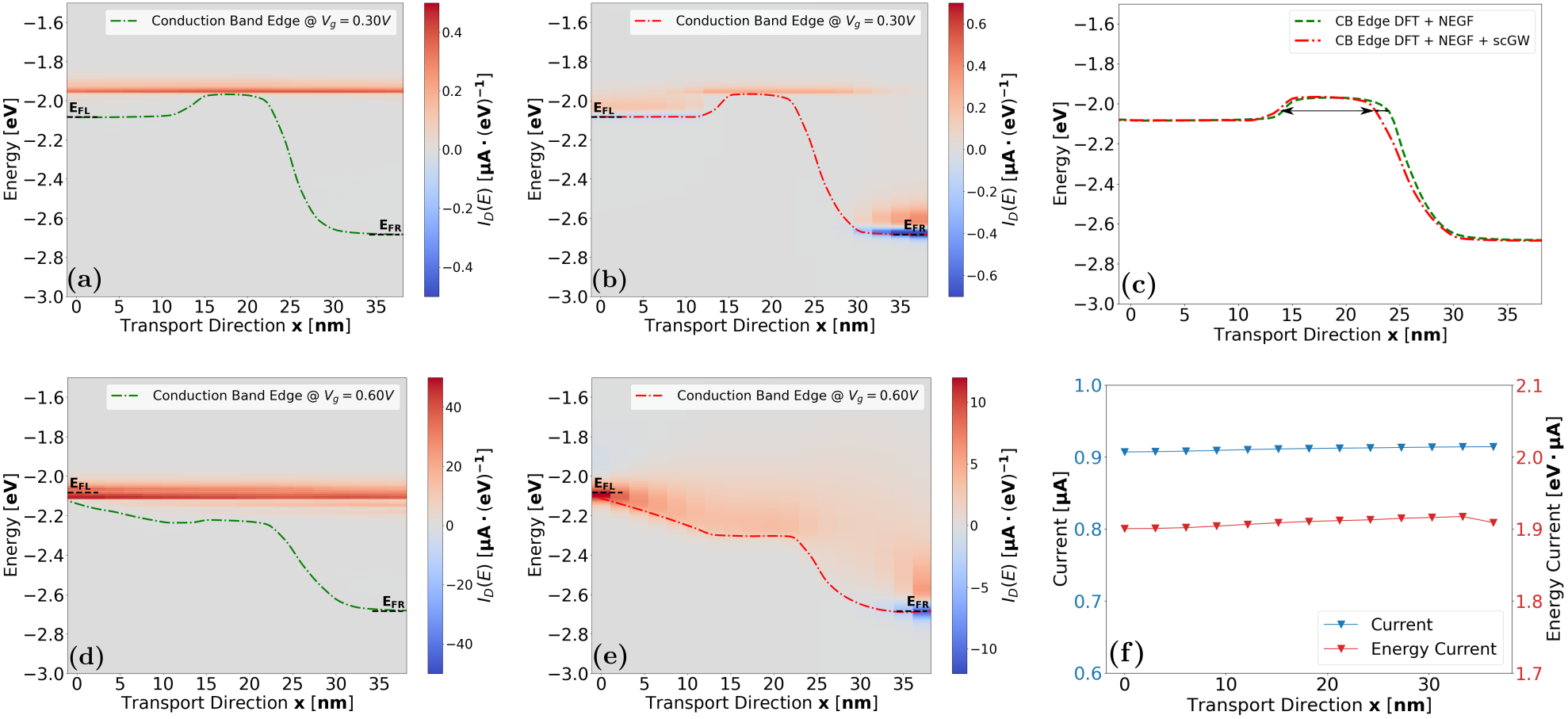}
\caption{\textbf{(a)} Energy- and position-resolved current flowing through the Si NWFET of Fig.~\ref{fig:nw_sketch} at $V_{GS}$=0.3 V, in the quasi-ballistic limit of transport. Red indicates high current concentrations from left to right. The dashed-dotted green light refers to the conduction band edge. The position of the source ($E_{FL}$) and drain ($E_{FR}$) Fermi levels is marked with dashed black lines. \textbf{(b)} Same as \textbf{(a)}, but in the presence of electron-electron interactions. In blue regions, the current direction is reversed (from right to left). \textbf{(c)} Conduction band edges from sub-plots \textbf{(a)} and \textbf{(b)} as a function of the transport direction $x$. The double arrow measures the effective length of the potential barrier. \textbf{(d)} Same as \textbf{(a)}, but at $V_{GS}$=0.6 V. \textbf{(e)} Same as \textbf{(b)}, but at $V_{GS}$=0.6 V. \textbf{(f)} Spatially-resolved electronic (blue line with triangles) and energy (red line with circles) currents computed with DFT+NEGF+scGW at $V_{GS}$=0.6 V. Both quantities are conserved throughout the device within a pre-defined relative tolerance of $\leq$1\%.}
\label{fig:SINW_spectral_current}
\end{figure*}

Current reduction is the most visible impact of electron-electron interactions, but by far not the only one. The spatial and energy distribution of the electrons propagating from source to drain is also significantly affected, as can be seen in Fig.~\ref{fig:SINW_spectral_current}. Sub-plots (a) and (b) present results at $V_{GS}$=0.3 V, in the subthreshold regime. Electron-electron interactions modify the current distribution on both sides of the potential barrier. On the left (source), scattering processes help electrons overcome the potential barrier separating them from the drain region. On the right (drain), they lead to a strong thermalization of the electron population. This phenomenon originates from the interaction of ``hot'' electrons traveling from the source of the transistor with those located in the drain at lower Fermi level. It is schematized in Fig.~\ref{fig:SINW_spectral_current}(b) where the solid dots represent electrons and the empty ones available states. The renormalization of the carrier population on the drain side of the NWFET is very similar to that of the SWCNT in Fig.~\ref{fig:spec_current_CNT_GW}(b). It is a clear signature of e-e interactions.

Since the spatial distribution of the electron population changes under the influence of scattering, the electrostatics does the same. Figure~\ref{fig:SINW_spectral_current}(c) compares the gate-modulated conduction band edge of the Si NWFET with and without scattering. It appears that the potential barrier separating the source from the drain is effectively longer in the quasi-ballistic case, roughly 10 vs. 8.5 nm when scattering is switched on. Hence, more source-to-drain tunneling leakage occurs in the presence of e-e interactions, which explains the slightly worse subthreshold slope of this configuration in the 0.25$\leq V_{GS}\leq$0.35 V range, namely 70 vs. 63 mV/dec.

The spectral current of another bias point ($V_{GS}$=0.6 V) is shown in Fig.~\ref{fig:SINW_spectral_current}(d) and (e). On the one hand, the quasi-ballistic current remains homogeneously distributed from source to drain and is lower-bounded by the source conduction band edge. On the other hand, the current with e-e scattering closely follows the conduction band edge till the drain side. There, the same thermalization effects as at lower voltages can be observed. Finally, note that the electronic and energy currents are conserved through the Si NWFET too, as demonstrated in Fig.~\ref{fig:SINW_spectral_current}(f). These currents do not vary by more than 1\% between source and drain, the convergence criterion that was set. From 400 to 500 self-consistent iterations between $\mathbf{G}\rightarrow\mathbf{P}\rightarrow\mathbf{W}\rightarrow\mathbf{\Sigma}$ are typically required to reach convergence. This achievement proves that the generalization of the Meier-Wingreen formula in Eqs.~(\ref{eq:meir-wingreen}) and (\ref{eq:energy_meir-wingreen}) works as expected, i.e., it satisfies current conservation.

\begin{figure*}[ht]
    \centering
        \includegraphics[width=1\linewidth]{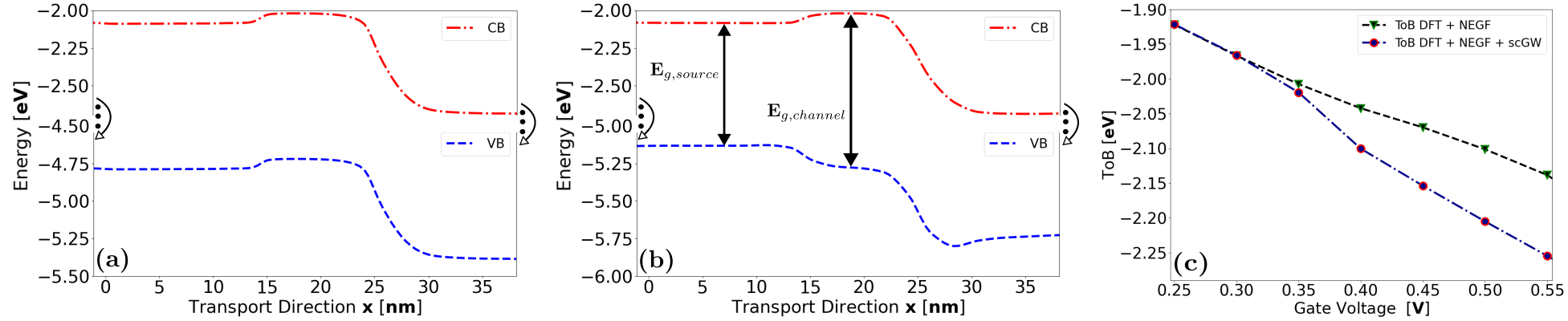}
              \caption{\textbf{(a)} Position-dependent conduction (red dashed-dotted line) and valence (blue dashed-dotted line) band edge of the Si NWFET from Fig.~\ref{fig:nw_sketch} at $V_{GS} = 0.35$ V and $V_{DS}$=0.6 V in the quasi-ballistic limit. The band gap remains constant and equal to 2.7 eV throughout the device. \textbf{(b)} Same as \textbf{(a)}, but with electron-electron scattering. The channel bandgap $E_{g, channel}$ is equal to 3.26 eV and is higher than in the source ($E_{g, source}$=3.05 eV). \textbf{(c)} Evolution of the conduction band edge at the ToB location as a function of the applied $V_{GS}$, in the quasi-ballistic case (green triangles) and in the presence of electron-electron interactions (red circles).}
            \label{fig:pos_bandgap}
\end{figure*}

By carefully inspecting the ``current vs. voltage'' characteristics in Fig.~\ref{fig:Id-Vg-curve}, it can be seen that the subthreshold slope in case of electron-electron scattering becomes steeper than its quasi-ballistic counterpart between $V_{GS}$=0.35 and 0.4 V. This is not a numerical artifact but a consequence of the fact that the band gap of semiconductors depends on the local carrier concentration. Band gap narrowing takes place when the free electron or hole population increases. This many-body effect is fully captured by the GW approximation through the real part of $\mathbf{\Sigma}^R_{GW}$ \cite{Abram1984}. As a result, the bandgap in the source and drain extensions, which host high electron concentrations (2.5e20 cm$^{-3}$), is smaller ($E_G$=3.05 eV) than in the channel region where there is initially almost no carrier ($E_G$=3.26 eV at $V_{GS}$=0.35 V). This maximum bandgap value, extracted at the top-of-the-barrier (ToB) location \cite{Rahman2003}, agrees well with the one computed in the equilibrium case, with the Fermi levels fixed in the middle of the gap ($E_G$=3.3 eV). Such bandgap modulations are absent from the quasi-ballistic simulations, they only manifest themselves when electron-electron scattering is accounted for, as depicted in Fig.~\ref{fig:pos_bandgap}(a) and (b).

However, the electron population in the channel increases with $V_{GS}$, thus causing a local decrease of the band gap. What matters in $n$-type transistors is the behavior of the conduction band edge at the top-of-the-barrier location. Its evolution as a function of $V_{GS}$ is reported in Fig.~\ref{fig:pos_bandgap}(c), both in the quasi-ballistic limit and with electron-electron interactions. Obviously, the ToB moves faster towards lower energies under the influence of electron-electron scattering because the carrier-induced bandgap decrease adds up to the effect of the gate-to-source voltage. In other words, the potential barrier separating the source and drain is more rapidly reduced when scattering is present, which explains the aforementioned boost in the current between $V_{GS}$=0.35 and 0.4 V.

Standard formula describing the evolution of the barrier height \cite{neamen2003semiconductor} or top-of-the-barrier energy \cite{Rahman2003} as a function of the gate voltage ignore the renormalization of the band gap as the carrier population increases. We believe that this effect plays an important role in highly confined transistor geometries with high electron and/or hole concentrations. We therefore propose to modify the existing equation governing the barrier height or ToB, labeled $U_B$, by adding a term accounting for the bandgap change $\Delta E_{g}$
\begin{align}
    U_B(V_{GS})\quad=\quad U_{B0}-\alpha\cdot q\cdot V_{GS} +  \nonumber \\
    Q_S(V_{GS})/C_{G}-\beta \cdot \Delta E_g(V_{GS}),
\label{eq:eg}    
\end{align}
where $U_{B0}$ is the barrier height at $V_{GS}$=0, $\alpha$ the lever arm of the gate electrode (0$\leq\alpha\leq$1), $q$ the elementary charge, $Q_S(V_{GS})$ the gate-induced electron charge in the channel, $C_G$ the gate capacitance, and $\Delta E_g(V_{GS})$ the positive band gap decrease caused by the growing electron population. The factor 0$\leq\beta\leq$1 refers to the fraction of $\Delta E_{g}$ affecting the conduction band. Overall, the band gap decrease ($-\beta \cdot \Delta E_g$) partly compensates for the reduction of the gate modulation efficiency ($-\alpha\cdot q\cdot V_{GS}$) caused by the charge $Q_S$, i.e., it counterbalances the term $+Q_S(V_{GS})/C_{G}$ in Eq.~(\ref{eq:eg}). Hence, it can be seen as a positive factor. Practically, previous studies reported the band gap narrowing $\Delta E_g$ of Si and other relevant semiconductors as a function of the electron and hole density, see for example \cite{Jain1991}. These values can be directly plugged into Eq.~(\ref{eq:eg}).

\section{Conclusions}\label{sec:Conclusions}
We have implemented an \textit{ab initio} atomistic quantum transport approach based on density functional theory, the non-equilibrium Green's function (NEGF) formalism, and the self-consistent GW approximation (DFT+NEGF+scGW). This many-body perturbation method allows to account for electron-electron interactions, which have been found to have a substantial impact on the performance of next-generation ultra-scaled transistors. By truncating the effective Coulomb interactions and leveraging the recursive Green's function algorithm to obtain the screened interaction $\mathbf{W}$, we have been able to treat systems made of up to $\approx$ 3000 atoms. These are probably the first ``large-scale'' quantum transport simulations including carrier-carrier scattering under strong non-equilibrium conditions.

Considering a single-wall carbon naotube as first example, we have established, on the one hand, that a large cut-off is needed for the electron-electron interactions to converge although the qualitative picture does not change much, regardless of the cut-off choice. Next, the transport properties of an ultra-scale Si gate-all-around nanowire field-effect transistor have been extensively analyzed, highlighting the role of electron-electron interactions on the current magnitude as well as on its energy and spatial distribution. We have uncovered another phenomenon relevant to the operation of transistors with highly confined carrier populations, namely the impact of the electron-induced narrowing of the Si band gap in the channel region. This effect could potentially offset the reduction of the gate modulation efficiency associated with increased channel carrier concentrations. Our DFT+NEGF+scGW method allows to investigate whether some materials and device configurations could possibly favor this compensation mechanism and boost the current of next-generation nano-transistors.

\section*{Acknowledgments}
This work received funding from the Swiss National Science Foundation (SNSF) under the grant agreement ``Quantum Transport Simulations at the Exascale and Beyond (QuaTrEx)'' ($\mathrm{n^\circ}$ 209358) and ``NCCR MARVEL'' ($\mathrm{n^\circ}$~205602). We acknowledge support from the Swiss National Supercomputing Center (CSCS) under projects s1119 and lp16 and from the Finish IT Center for Science (CSC) under project 465000929. J.~C. acknowledges funding from the European Union under the Marie Skłodowska-Curie grant No. 885893.

\FloatBarrier
\appendix

\section{Derivation of the GW self-energy equations in the energy domain}\label{app:gw}
Following the derivation of \cite{PhysRevB.77.115333}, the GW equations in the time domain are given by
\begin{align}
    \label{sigma_contour}
    \Sigma_{GW,ij}(\tau, \tau')  &= iG_{ij}(\tau, {\tau'}^{+})W_{ij}(\tau, \tau')
    \\
    \label{w_contour}
     W_{ij}(\tau, \tau') &= \widetilde{V}_{ij}\delta_{C}(\tau, \tau') \nonumber
     \\
                         + \sum_{kl} \int_{C} &d\tau_1 \widetilde{V}_{ik}P_{kl}(\tau, \tau_1)W_{lj}(\tau_1, \tau')
    \\
    P_{ij}(\tau, \tau') &= -iG_{ij}(\tau, \tau')G_{ji}(\tau', \tau),
\end{align}
where $\tau$, $\tau_1$, and $\tau'$ are times situated on a complex contour. We can apply Langreth-rules \cite{2008} to the self-energy (Eq.~\eqref{sigma_contour}) to obtain its lesser/greater components. Furthermore, since we are interested in steady-state, the tuple $(\tau, \tau')$ can be replaced with $t=\tau-\tau'$. These modifications lead to the following equation in the time domain for $\Sigma^{\lessgtr}_{GW,ij}(t)$:
\begin{equation}
\label{eq:siglgtr_timedomain}
\Sigma^{\lessgtr}_{GW,ij}(t) = iG^{\lessgtr}_{ij}(t)W^{\lessgtr}_{ij}(t),
\end{equation}
which can be seamlessly transformed into the energy domain, giving rise to Eq.~\eqref{eq:siglg}.
The time-domain equation for the retarded self-energy is more complicated to handle as it is made of two parts, an exchange term, $\Sigma^{R}_{x,ij}(t)$, containing a delta function and a correlation term, $\Sigma^{R}_{corr,ij}(t)$,
\begin{equation}
\label{eq:sigr_timedomain}
\Sigma^{R}_{GW,ij}(t) = \Sigma^{R}_{x,ij}(t)+\Sigma^{R}_{corr,ij}(t).
\end{equation}
While $\Sigma^{R}_{x,ij}(t)=iG^{<}_{ij}(t=0)\widetilde{V}_{ij}$ can be directly derived by combining Eqs.~(\ref{sigma_contour}) and (\ref{w_contour}), $\Sigma^{R}_{corr,ij}(t)$ must be carefully handled. It is defined as
\begin{align}
\label{sigma_corr_timedomain}
\Sigma^{R}_{corr,ij}(t) & = \theta(t) \cdot \lbrack \Sigma^{>}_{GW,ij}(t) - \Sigma^{<}_{GW,ij}(t), \rbrack
\end{align}
which, by replacing $\Sigma^{\lessgtr}_{GW,ij}(t)$ with its definition, can be written in three different ways
\begin{eqnarray}
\label{sigma_corr_1}
\Sigma^{R}_{corr,ij}(t) & = & G^{>}_{ij}(t)W^{R}_{corr,ij}(t)+G^{R}_{ij}(t)W^{<}_{ij}(t),\\
\label{sigma_corr_2}
& = & G^{<}_{ij}(t)W^{R}_{corr,ij}(t)+G^{R}_{ij}(t)W^{>}_{ij}(t),\\
\label{sigma_corr_3}
& = &G^{R}_{ij}(t)W^{R}_{corr,ij}(t)+G^{R}_{ij}(t)W^{<}_{ij}(t)+\\\nonumber
&& G^{<}_{ij}(t)W^{R}_{corr,ij}(t),
\end{eqnarray}
with $W^{R}_{corr} = \theta(t)\cdot \lbrack W^{>}(t) - W^{<}(t) \rbrack$. All these expressions for $\Sigma^{R}_{corr,ij}(t)$ are equivalent in the time domain. However, in the energy domain, they have different values and none of them appears to be correct. For example, and as expected, the imaginary part of Eqs.~(\ref{sigma_corr_1}) and (\ref{sigma_corr_2}) in the energy domain is equal to $(\Sigma^{>}_{GW}(E)-\Sigma^{<}_{GW}(E))/2$, but this is not the case of Eq.~(\ref{sigma_corr_3}). Similarly, the real part of Eq.~(\ref{sigma_corr_3}) is the only one that produces the correct band gap behavior, i.e., $E_G$ increase when electron-electron interactions are taken into account.

Since Eqs~(\ref{sigma_corr_1}) to (\ref{sigma_corr_3}) fail at delivering accurate results, we opted for another solution, namely directly transforming Eq.~(\ref{sigma_corr_timedomain}) into the energy domain, which yields \cite{Lake1997}
\begin{equation}
\Sigma^{R}_{corr, ij}(E) = - \frac{i}{2}\Gamma_{ij}(E) + \mathcal{P} \text{ } \int_{E} \frac{dE'}{2\pi}\frac{\Gamma_{ij}(E')}{E-E'}.
\end{equation}
Here, $\mathcal{P}$ denotes the principal part of the integral and $\Gamma_{ij}$ is the broadening function
\begin{equation}
    \Gamma_{ij} = i \lbrack \Sigma^{>}_{GW,ij} - \Sigma^{<}_{GW,ij} \rbrack.
\end{equation}
In \cite{PhysRevB.77.115333} it was proposed to first compute $\Sigma^{\lessgtr}_{GW}(E)$ in the energy domain, then fast Fourier transform (FFT) these quantities to obtain their time-dependent counterparts, apply Eq.~(\ref{sigma_corr_timedomain}) by removing all entries corresponding to $t<0$, and finally perform an inverse FFT to get $\Sigma^{R}_{corr}(E)$. We tested this approach as well, but found that the retarded self-energy it returns is not correct, neither its real nor its imaginary part.

\section{Extension of the RGF algorithm to $\mathbf{W}$}\label{app:RGF_new}
To capture long-range correlation effects, off-diagonal blocks of $\mathbf{G}$, $\mathbf{\Sigma}$, $\mathbf{P}$, and $\mathbf{W}$ are needed. Here, we assume that the matrix elements $\Sigma_{ij}$, $\widetilde{V}_{ij}$, and $P_{ij}$ decrease sufficiently rapidly to neglect them beyond a certain cut-off distance $r_{max}$. Therefore, the matrices $\left(E-\mathbf{H}-\mathbf{V_{ext}}-\mathbf{\Sigma}^{R}_{B}(E)- \mathbf{\Sigma}^{R}_{GW}(E)\right)$ in Eqs.~\eqref{eq:GR} and $\mathbf{M}^R$ in Eq.~\eqref{eq:WR} exhibit a block-tridiagonal structure, provided that the block size $N_{BS}$ is chosen large enough. Moreover, to compute the $\Sigma_{ij}$ and $P_{ij}$ entries satisfying $|x_i-x_j|\leq r_{max}$, only the corresponding elements of the $\mathbf{G}$ and $\mathbf{W}$ matrices are needed, as can be inferred from Eqs.~(\ref{eq:siglg}) and (\ref{eq:Plessgtr}).

The recursive Green's function (RGF) algorithm was specifically developed to produce selected entries of the inverse of a block-tridiagonal matrix and its associated lesser and greater components \cite{rgf}. It can, however, not be directly applied to the calculation of the screened Coulomb interaction $\mathbf{W}$ because of the presence of the bare Coulomb matrix $\mathbf{\widetilde{V}}$ on the right-hand-side of Eq.~(\ref{eq:WR}), instead of the identity matrix $\mathbf{I}$. This is why we introduced the quantity $\mathbf{X}^R$, which is defined as $\mathbf{M}^R\cdot\mathbf{X}^R=\mathbf{I}$. It can be used to reconstruct $\mathbf{W}^R=\mathbf{X}^R\cdot\mathbf{\widetilde{V}}$. Our approach allows to apply the same backward and forward recursion formulas as for $\mathbf{G}^R$ to compute $\mathbf{X}^{R}$. The forward recursion for the right-connected $\mathbf{x}^R$ is given by \cite{rgf}
\begin{equation}
    \mathbf{x}^R_{n, n} = \lbrack \mathbf{M}_{n, n}  + \mathbf{M}_{n, n+1} \mathbf{x}^{R}_{n+1,n+1} \mathbf{M}_{n+1, n}  \rbrack^{-1}
\end{equation}
with $\mathbf{x}^R_{N,N}=\mathbf{M}_{N,N}^{-1}$. During the backward pass, both the diagonal blocks of the full $\mathbf{X}^R$ matrix
\begin{align}
    \mathbf{X}^R_{n, n} & = \mathbf{x}^R_{n, n} +
    \mathbf{x}^R_{n, n}\lbrack \mathbf{M}_{n, n-1} \mathbf{X}^{R}_{n-1, n-1} \mathbf{M}_{n-1, n}  \rbrack \mathbf{x}^R_{n, n},
\end{align}
and the off-diagonal ones
\begin{align}
    \mathbf{X}^{R}_{n, n+1} &= -\mathbf{X}^R_{n, n} \mathbf{M}_{n, n+1} \mathbf{x}^{R}_{n+1, n+1}\\
    \mathbf{X}^{R}_{n+1, n} &= -\mathbf{x}^{R}_{n+1, n+1} \mathbf{M}_{n+1, n} \mathbf{X}^R_{n, n}
\end{align}
can be calculated. This recursion starts with $\mathbf{X}^R_{1,1}=\mathbf{x}^R_{1,1}$. Although $\mathbf{W}^R$ is not needed, its diagonal blocks can be evaluated based on the available $\mathbf{X}^R$, i.e., $\mathbf{W}^R_{n,n}=\sum_{m}\mathbf{X}^R_{nm}\cdot\mathbf{\widetilde{V}}_{mn}$, where $m=n$ or $n\pm1$. It should be noted that the first and last blocks of $\mathbf{\widetilde{V}}$ must be modified to account for the open nature of the simulated systems. According to Eq.~(\ref{eq:sigb}), the boundary blocks $\mathbf{S}^R_{B(1,1)}$ and $\mathbf{S}^R_{B(N,N)}$ in Eq.~(\ref{eq:WR}) have the following form $\mathbf{S}^R_{B(i,i)}=\mathbf{A}_{i,j}\cdot \mathbf{x}^R_C\cdot\mathbf{A}_{i,j}$. If open boundary conditions are included, $\mathbf{\widetilde{V}}_{i,i}$ becomes $\mathbf{\widetilde{V}}_{i,i}-\mathbf{M}_{i,j}\cdot\mathbf{x}^R_C\cdot\mathbf{\widetilde{V}}_{j,i}$ with $(i,j)$ equal to $(1,0)$ or $(N,N+1)$, while $C$ represents the left or right contact.

Solving for $\mathbf{X}^R$ instead of $\mathbf{W}^R$ directly affects the treatment of $\mathbf{W^{\lessgtr}}$. To keep it similar to $\mathbf{G}^{\lessgtr}$ in Eq.~\eqref{eq:lessgtrGLGtilda} we defined $\mathbf{L}^{\lessgtr}=\mathbf{\widetilde{V}}\cdot\mathbf{P}^{\lessgtr}\cdot\mathbf{\widetilde{V}}^T$ in Eq.~\eqref{eq:WLG}. Therefore, the same recursion relations as for $\mathbf{G}^{\lessgtr}$ can be applied to $\mathbf{W}^{\lessgtr}$. For the right-connected $\mathbf{w}^{\lessgtr}$, the forward pass of the RGF algorithm reads
\begin{align}
\label{eq:wleftconn}
    \mathbf{w}^{\lessgtr}_{n, n} =& \mathbf{x}^R_{n, n} \lbrack \mathbf{L}^{\lessgtr}_{n, n} + \mathbf{M}_{n, n+1} \mathbf{w}^{\lessgtr}_{n+1,n+1} {\mathbf{M}^{\dagger}_{n,n+1}} - \nonumber \\ 
     &\left( \mathbf{a}^{\lessgtr}_{n, n} - 
      (\mathbf{a}^{\lessgtr}_{n, n})^{\dagger} \right) 
     \rbrack (\mathbf{x}^{R}_{n, n}) ^{\dagger},
\end{align}
where
\begin{equation}
 \mathbf{a}^{\lessgtr}_{n, n} = \mathbf{M}_{n, n+1} \mathbf{x}^R_{n+1, n+1} \mathbf{L}^{\lessgtr}_{n+1, n}
\end{equation}
and
\begin{equation}
 \mathbf{w}^{\lessgtr}_{N, N} = \mathbf{x}^R_{N, N}\cdot \lbrack\mathbf{L}^{\lessgtr}_{N, N}+\mathbf{S}_{B(N,N)}^{\lessgtr}\rbrack\cdot (\mathbf{x}^{R}_{N,N}) ^{\dagger}. 
\end{equation}
The backward pass yields the full $\mathbf{W}^{\lessgtr}$ matrix. We start with the diagonal blocks
\begin{align}
     \mathbf{W}^{\lessgtr}_{n, n}  &= \mathbf{w}^{\lessgtr}_{n,n} +
    \mathbf{x}^R_{n, n} \lbrack \mathbf{M}_{n, n-1} \mathbf{W}^{\lessgtr}_{n-1,n-1} \mathbf{M}^{\dagger}_{n, n-1} \rbrack (\mathbf{x}^R_{n, n}) ^{\dagger} - \nonumber \\
    & \left( \mathbf{A}^{\lessgtr}_{n, n} - (\mathbf{A}^{\lessgtr}_{n, n})^{\dagger} \right) + \left( \mathbf{B}^{\lessgtr}_{n, n} - (\mathbf{B}^{\lessgtr}_{n, n})^{\dagger} \right),
\end{align}
where
\begin{align}
    \mathbf{A}^{\lessgtr}_{n, n} &= \mathbf{x}^R_{n, n} \mathbf{L}^{\lessgtr}_{n, n-1} \lbrack \mathbf{x}^R_{n, n}  \mathbf{M}_{n, n-1} \mathbf{X}^{R}_{n-1, n-1} \rbrack ^{\dagger},\\
    \mathbf{B}^{\lessgtr}_{n, n} &= \mathbf{x}^R_{n, n}  \mathbf{M}_{n, n-1} \mathbf{X}^{R}_{n-1, n-1}  \mathbf{M}_{n-1, n} \mathbf{w}^{\lessgtr}_{n, n},\\
    \mathbf{W}^{\lessgtr}_{1,1}&=\mathbf{w}^{\lessgtr}_{1,1}.
\end{align}
The off-diagonal blocks can be assembled from
\begin{align}
\mathbf{W}^{\lessgtr}_{n,n+1}= &\mathbf{X}^{R}_{n, n} \lbrack \mathbf{L}^{\lessgtr}_{n, n+1} \left({\mathbf{x}^{R}_{n+1, n+1}}\right)^{\dagger} - \nonumber\\
&\mathbf{M}_{n, n+1} \mathbf{w}^{\lessgtr}_{n+1,n+1}\rbrack - \nonumber\\
&\mathbf{W}^{\lessgtr}_{n, n} \left( \mathbf{x}^R_{n+1, n+1}  \mathbf{M}_{n+1, n} \right) ^{\dagger}.
\end{align}
As an important side note, we would like to stress out that Eqs.~\eqref{eq:WR} and \eqref{eq:WLG} contain products of two $\mathbf{K}^{R} = \mathbf{\widetilde{V}}\cdot\mathbf{P}^R$ and three $\mathbf{L}^{\lessgtr} = \mathbf{\widetilde{V}}\cdot\mathbf{P}^{\gtrless}\cdot\mathbf{\widetilde{V}^{T}}$ block-tridiagonal matrices, respectively. These operations increase the bandwidth of $\mathbf{K}^{R}$ (penta-diagonal) and $\mathbf{L}^{\lessgtr}$ (hepta-diagonal). To still be able to use the RGF algorithm, the size of the involved blocks must be increased to make the resulting matrices block-tridiagonal. Also, if the left (right) contact is assumed to be a repetition of the first (last) unit cell of the device, the first (last) diagonal blocks of $\mathbf{K}^{R}$ and $\mathbf{L}^{\lessgtr}$ must be slightly modified by adding the following terms to them
\begin{align}
    \mathbf{dK}^{R}_{i, i} &= \mathbf{\widetilde{V}}_{i, j} \mathbf{P}^{R}_{j, i} \\
    \mathbf{dL}^{\lessgtr}_{i ,i} &= \mathbf{\widetilde{V}}_{i, j} \mathbf{P^{\lessgtr}}_{j, j}\mathbf{\widetilde{V}}_{j, i} + \mathbf{\widetilde{V}}_{i, j}\mathbf{P^{\lessgtr}}_{j, i}\mathbf{\widetilde{V}}_{i, i} + \mathbf{\widetilde{V}}_{i, i}\mathbf{P^{\lessgtr}}_{i, j}\mathbf{\widetilde{V}}_{j, i}.
\end{align}
Here, again $(i,j)$ corresponds to $(1,0)$ for the left contact and $(N,N+1)$ for the right contact.

\section{Generalization of the Meir-Wingreen formula}\label{app:current}
To compute the current flowing through open systems, the Meir-Wingreen formula is very convenient. It returns the value from a lead $B$ into the first unit cell attached to it \cite{Ridley2022}
\begin{align}
\label{eq:meir-wingreen_orig}
        I_{n} = -\frac{2q}{\hbar} \mathfrak{Re} \int_{-\infty} ^ {\infty}  \frac{dE}{2\pi} \text{ } \mathrm{tr}& \{ \mathbf{\widetilde{\Sigma}}^{>}_{B,n}(E)\mathbf{G}^{<}_{n,n}(E) - \nonumber \\ 
        &  \mathbf{G}^{>}_{n,n}(E)\mathbf{\widetilde{\Sigma}}^{<}_{B,n}(E) \}.
\end{align}
Here,  $\mathbf{G}^{\lessgtr}_{n,n}$ is the $n^{\mathrm{th}}$ diagonal block of the $\mathbf{G}^{\lessgtr}$ matrix, $\mathbf{\widetilde{\Sigma}}^{\lessgtr}_{B,n}$ is the boundary self-energy corresponding to lead $B$, and the index $n$ is either equal to 1 or $N$ in devices with two contacts (source on the left, drain on the right).

To generalize Eq.~(\ref{eq:meir-wingreen_orig}) we draw inspiration from Eq.~(\ref{eq:wleftconn}), which established a recursion to calculate the right-connected block $\mathbf{w}^{\lessgtr}_{n,n}$ given$\mathbf{w}^{\lessgtr}_{n+1,n+1}$. In fact, Eq.~(\ref{eq:wleftconn}) propagates the influence of the right boundary self-energy, $\mathbf{S}_{B(N,N)}^{\lessgtr}$, from its original location at unit cell $N$ to any position $n$. The same type of relationship exists for the right-connected electron Green's function $\mathbf{g}^{\lessgtr}_{n,n}$ \cite{Lake1997}. It can, thus, be used to push the right boundary to any unit cell inside the device with index $1\leq n\leq N-1$. Hence, a novel boundary self-energy $\mathbf{\widetilde{\Sigma}}^{\lessgtr}_{B,n}$ can be defined
\begin{equation}
\label{eq:sigmab_mw}
\mathbf{\widetilde{\Sigma}}^{\lessgtr}_{B,n}=\mathbf{M}_{n, n+1} \mathbf{g}^{\lessgtr}_{n+1,n+1} {\mathbf{M}^{\dagger}_{n,n+1}} - \left( \mathbf{a}^{\lessgtr}_{n, n} - 
      (\mathbf{a}^{\lessgtr}_{n, n})^{\dagger} \right) 
\end{equation}
with
\begin{equation}
\mathbf{M}=\left(E-\mathbf{H}-\mathbf{V_{ext}}-\mathbf{\Sigma}^{R}_{GW}(E)\right)
\end{equation}
and
\begin{equation}
\mathbf{a}^{\lessgtr}_{n, n} = \mathbf{M}_{n, n+1} \mathbf{g}^R_{n+1, n+1} \mathbf{\Sigma}^{\lessgtr}_{n+1, n}.
\end{equation}
Here, $\mathbf{g}^R$ is the right-connected retarded Green's function. Taking advantage of Eq.~(\ref{eq:sigmab_mw}), the Meir-Wingreen formula in Eq.~(\ref{eq:meir-wingreen_orig}) can be generalized to any unit cell index $n$: The matrix $\mathbf{\Sigma}^{\lessgtr}_{B,n}$ is ``simply'' replaced by $\mathbf{\widetilde{\Sigma}}^{\lessgtr}_{B,n}$.

Practically, we recall the traditional Meir-Wingreen equation for unit cells $n$=1 and $n$=$N$ and apply its generalized form to $2\leq n\leq N-1$. As can be seen in the energy- and position-resolved currents in Figs.~\ref{fig:spectral_currents_CNT} and \ref{fig:SINW_spectral_current}, the current distribution does not abruptly change at the $n=1\rightarrow 2$ and $n=N-1\rightarrow N$ interfaces. Also, both the electronic and energy currents are conserved. These observations underline the validity of the generalized Meir-Wingreen formula.

\section{Screened Interaction with Open Boundary Conditions}\label{app:obcw}
We included open boundary conditions (OBCs) in the calculation of both the retarded and lesser/greater screened Coulomb interactions in Eqs.~\eqref{eq:WR} and \eqref{eq:WLG}. These OBCs are cast into the $\mathbf{S}^{R}_{B}(E)$ and $\mathbf{S}^{\lessgtr}_{B}(E)$ matrices, which have the same sparsity pattern as their electronic counterparts $\mathbf{\Sigma}^{R}_{B}(E)$ and $\mathbf{\Sigma}^{\lessgtr}_{B}(E)$. In structures with two contacts, one on the left and one on the right of the central device part, as the considered single-wall carbon nanotube and silicon nanowire, only the upper left and lower right corners of $\mathbf{S}^{R}_{B}(E)$ and $\mathbf{S}^{\lessgtr}_{B}(E)$ are different from zero. The computation of these blocks was discussed in Section \ref{subsec:boundaries}.

\begin{figure*}[!t]
\centering
\includegraphics[width=\linewidth]{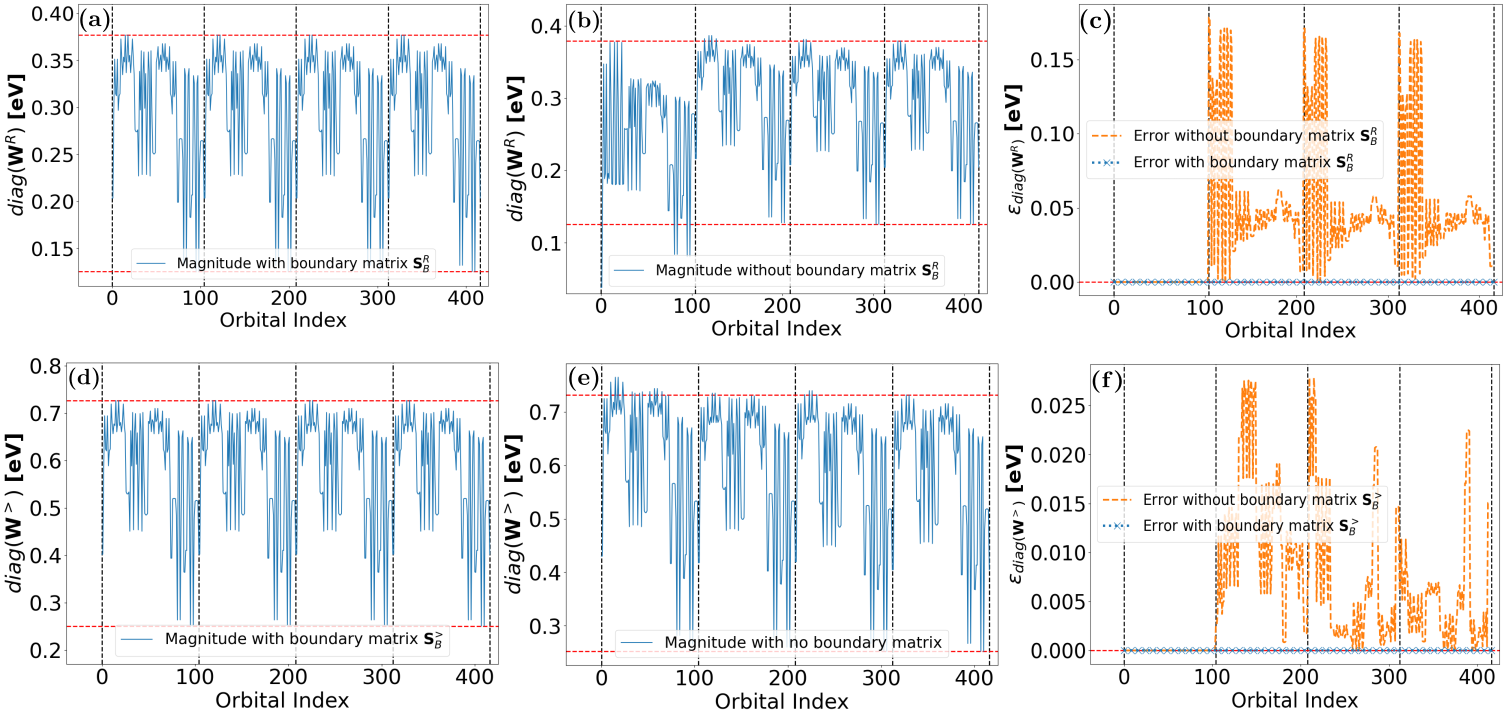}
\caption{Influence of the open boundary conditions (OBC) on the behavior of the diagonal entries of the screened Coulomb matrix $\mathbf{W}$ for the Si nanowire from Fig.~\ref{fig:nw_sketch} under flat-band conditions. The first 416 entries corresponding to the first four unit cells of $\mathbf{W}$ are shown for one energy point. There are $N_O$=104 orbitals per cell. \textbf{(a)} $\mathbf{W}^R$ with its OBC matrix $\mathbf{S}^{R}_{B}$. \textbf{(b)} $\mathbf{W}^R$ without $\mathbf{S}^{R}_{B}$. \textbf{(c)} Error in the $\mathbf{W}^R$ entries when $\mathbf{S}^{R}_{B}$ is included (dotted blue line) and when it is not (dashed orange line). The error is measured for each unit cell with respect to the results of the first unit cell, i.e., the error is defined as $\textrm{diag}(\mathbf{W}^R(N_O*(n-1)+1:N_O*n,N_O*(n-1)+1:N_O*n)-\mathbf{W}^R(1:N_O,1:N_O))$, where $n$ goes from 1 to 4. \textbf{(d)} $\mathbf{W}^>$ with its OBC matrix $\mathbf{S}^{>}_{B}$. \textbf{(e)} $\mathbf{W}^>$ without $\mathbf{S}^{>}_{B}$. \textbf{(f)} Same as \textbf{(c)}, but for $\mathbf{W}^>$.}
\label{fig:W_boundary_comparison}
\end{figure*}

Figure~\ref{fig:W_boundary_comparison} demonstrates the necessity of incorporating $\mathbf{S}^{R}_{B}(E)$ and $\mathbf{S}^{\lessgtr}_{B}(E)$. There, the diagonal entries of $\mathbf{W}^R$ and $\mathbf{W}^{>}$ are reported for the first four unit cells of the Si nanowire from Fig.~\ref{fig:nw_sketch}, at one given energy point, under flat-band conditions. Each unit cell contains 104 orbitals. Because of the imposed periodicity (the left (right) contact is made of the repetition of the first (last) unit cell of the device and all device unit cells are identical under flat-band conditions), the diagonal entries of $\mathbf{W}^R$ and $\mathbf{W}^{>}$ shown in Fig.~\ref{fig:W_boundary_comparison} should only vary within each cell, but not from one cell to the other. To verify whether this property is satisfied, we computed the error between each cell and the first one. It is  plotted in Fig.~\ref{fig:W_boundary_comparison}(c) and (f) for $\mathbf{W}^R$ and $\mathbf{W}^{>}$, respectively, with and without the OBCs. It can be clearly observed that the results for all unit cells are identical to each other, as expected, only when the OBCs are taken into account. Hence, their presence is not only justified, but required. Similar results are found if the last four unit cells of the Si nanowire are represented instead of the first four ones.

\section{Solution of the Lyapunov-type equation for $\mathbf{S}^{\lessgtr}_B$}\label{app:obcw}
In Section \ref{subsec:boundaries}, it was established that the blocks of the open boundary matrix $\mathbf{S}^{\lessgtr}_B$ depend on the knowledge of $\mathbf{w}^{\lessgtr}_{i,i}$, which obeys the following equation 
\begin{align}
 \mathbf{w}^{\lessgtr}_{i, i} &= \mathbf{x}^R_{i, i} \lbrack \mathbf{L}^{\lessgtr}_{i, i} + \mathbf{M}_{i, j} \mathbf{w}^{\lessgtr}_{i,i} {\mathbf{M}^{\dagger}_{i, j}} - 
    \left( \mathbf{a}^{\lessgtr}_{i, i} - {\mathbf{a}^{\lessgtr}_{i, i}}^{\dagger} \right)
       \rbrack (\mathbf{x}^R_{i, i}) ^{\dagger}.
\end{align}
This equation can be re-written as
\begin{equation}
\label{eq:Dis_lyapunov}
\mathbf{w}^{\lessgtr}_{i, i} - \mathbf{\Pi} \mathbf{w}^{\lessgtr}_{i, i} \mathbf{\Pi}^{\dagger} = \mathbf{\Omega},
\end{equation}
where
\begin{equation}
\mathbf{\Pi} = \mathbf{x}^R_{i, i} \mathbf{M}_{i,j}
\end{equation}
and
\begin{equation}
\mathbf{\Omega} = \mathbf{x}^R_{i, i} \lbrack L^{\lessgtr}_{i, i} - \left( \mathbf{a}^{\lessgtr}_{i, i} - {\mathbf{a}^{\lessgtr}_{i, i}}^{\dagger} \right)  \rbrack (\mathbf{x}^R_{i, i}) ^{\dagger}.
\end{equation}
Eq.~\eqref{eq:Dis_lyapunov} constitutes a Lyapunov discrete equation. It can be solved by transforming it into an equivalent system with the help of the eigenvectors and eigenvalues of $\mathbf{\Pi}$
\begin{equation}
    \mathbf{\Pi} = \mathbf{U}\mathbf{\Lambda_{\Pi}}\mathbf{U}^{-1}.
\end{equation}
The eigenvalues $\lambda_i$ of $\mathbf{\Pi}$ sit on the diagonal of $\mathbf{\Lambda_{\Pi}}$.
Multiplying Eq.~\eqref{eq:Dis_lyapunov} from the left with $\mathbf{U}^{-1}$ and from the right with $({\mathbf{U}^{-1}})^{\dagger}$ transforms the original system into
\begin{equation}
 \mathbf{\widetilde{w}}^{\lessgtr} - \mathbf{\Lambda_{\Pi}} \mathbf{\widetilde{w}}^{\lessgtr}\mathbf{\Lambda_{\Pi}} ^{\dagger} = \mathbf{\widetilde{\Omega}}
\end{equation}
where $\mathbf{\widetilde{w}}^{\lessgtr}=\mathbf{U}^{-1}\mathbf{w}^{\lessgtr}_{i,i}({\mathbf{U}^{-1}})^{\dagger}$ and $\mathbf{\widetilde{\Omega}}=\mathbf{U}^{-1}\mathbf{\Omega}({\mathbf{U}^{-1}})^{\dagger}$. Because $\mathbf{\Lambda_{\Pi}}$ is diagonal, the entries of $\mathbf{\widetilde{w}}$ can be directly computed from
\begin{equation}
\widetilde{w}_{i, j} = \frac{1}{1-\lambda_i {\lambda_j}^{*}} \widetilde{\Omega}_{i,j}.
\end{equation}
Finally, $\mathbf{w}^{\lessgtr}_{i, i}$ can be reconstructed from $\mathbf{\widetilde{w}}^{\lessgtr}$
\begin{equation}
\mathbf{w}^{\lessgtr}_{i, i}  = \mathbf{U} \mathbf{\widetilde{w}}^{\lessgtr} \mathbf{U}^{\dagger}.
\end{equation}
This approach allows to compute both the left- and right-connected $\mathbf{w}^{\lessgtr}_{i, i}$, i.e., $i$=1 or $N$. Practically, it was found that several eigenvalues of $\mathbf{\Pi}$ are very small and can be excluded. By doing so, the size of the system of equations to be solved can be greatly reduced.

\FloatBarrier
\bibliography{apssamp}
\end{document}